\documentclass[fleqn,usenatbib]{mnras}

\usepackage{newtxtext,newtxmath}

\usepackage[T1]{fontenc}

\DeclareRobustCommand{\VAN}[3]{#2}
\let\VANthebibliography\thebibliography
\def\thebibliography{\DeclareRobustCommand{\VAN}[3]{##3}\VANthebibliography}

\usepackage{graphicx}	
\usepackage{amsmath}	
\usepackage{xspace}

\newcommand{\Q}{\, \mathbf{Q}}
\newcommand{\Qdot}{\, \dot{\mathbf{Q}}}

\usepackage{xparse}

\NewDocumentCommand{\Qdrag}{o}{%
  \, \dot{\mathbf{Q}}_{{\rm drag}\IfValueT{#1}{,#1}}%
}
\NewDocumentCommand{\Qmix}{o}{%
  \, \dot{\mathbf{Q}}_{{\rm mix}\IfValueT{#1}{,#1}}%
}
\NewDocumentCommand{\Qgrow}{o}{%
  \, \dot{\mathbf{Q}}_{{\rm grow}\IfValueT{#1}{,#1}}%
}

\newcommand{\model}{MOGLI\xspace}

\title[\model]{\model: Model for Multiphase Gas using Multifluid hydrodynamics}

\author[Das et al.]{
Hitesh Kishore Das,$^{1}$\thanks{E-mail: hitesh@mpa-garching.mpg.de}
Max Gronke,$^{1}$
Rainer Weinberger$^{2}$
\\
$^{1}$ Max Planck Institute for Astrophysics, Garching D-85748, Germany\\
$^{2}$ Leibniz-Institut f\"ur Astrophysik Potsdam (AIP), An der Sternwarte 16, 14482 Potsdam, Germany
}

\date{Draft from \today}

\pubyear{2024}

\begin{document}
\label{firstpage}
\pagerange{\pageref{firstpage}--\pageref{lastpage}}
\maketitle

\begin{abstract}
Multiphase gas, with hot ($\sim10^6$K) and cold ($\sim10^4$K) gas, is ubiquitous in astrophysical media across a wide range of scales. However, simulating multiphase gas has been a long-standing challenge, due to the large separation between the size of cold gas structures and the scales at which such gas impacts the evolution of associated systems. In this study, we introduce a new subgrid framework for such multiphase gas, \textbf{MOGLI}: Model for Multiphase Gas using Multifluid hydrodynamics, in multifluid \texttt{AREPO}. We develop this approach based on first principles and theoretical results from previous studies with resolved small-scale simulations, leading to a minimal number of free parameters in the formulation. We divide the interactions in the model into three sources: drag, turbulent mixing and cold gas growth.
As part of the model, we also include two methods for estimating the local turbulent velocities, one using the Kolmogorov scaling, and the other using the local velocity gradients.
We verify the different components of the framework through extensive comparison with benchmark single-fluid simulations across different simulation parameters, such as how resolved the cold gas is initially, the turbulent Mach number, spatial resolution, and random initialisation of turbulence. We test the complete scheme and a reduced version, with and without cold gas growth.
We find a very good qualitative and quantitative agreement across the different simulation parameters and diagnostics for both local turbulent velocity estimation methods. We also reproduce behaviour like the cold gas survival criteria as an emergent property. 
We discuss the applications and possible extensions of MOGLI and demonstrate its capability by running a simulation which would be computationally prohibitive to run as a resolved single-fluid simulation.
\end{abstract}


\section{Introduction}
Multiphase gas dynamics is prevalent in various astrophysical media, from accretion disks to galactic outflows. There are observational \citep{Tumlinson2017TheMedium,Veilleux2020}, numerical and theoretical \citep{McKee1977ASubstrate, Donahue2022, Faucher2023} evidence for the multiphase nature of astrophysical media. The multiphase nature of the interstellar medium (ISM), circumgalactic medium (CGM) and intracluster medium (ICM) significantly affects the evolution of corresponding systems, like galaxies or galaxy clusters, via processes such as their baryon cycles or feedback processes \citep{Veilleux2005GalacticWinds, Peroux2020}.

Recent observations have provided crucial insights into the structure of this multiphase gas, particularly in galactic halos. Studies have constrained the size of cold gas clouds in the CGM to be $\lesssim$ 10 pc \citep{Lan2017ApJ...850..156L, Crighton2015Metal-enrichedGalaxy, Schaye2007MNRAS.379.1169S, Rauch1999ApJ...515..500R, Chen2023ApJ...955L..25C}. This small-scale structure exists within large halos spanning tens to hundreds of kpc in radius \citep{Tumlinson2017TheMedium}, creating a significant challenge for numerical simulations attempting to resolve these disparate scales -- and leading to non-convergence of cosmological simulations in the halo gas content \citep[e.g.][]{Hummels2018, vandeVoort2021}.

Many theoretical studies use small-scale idealized simulation setups to better resolve and understand the small-scale gas structure. Some studies have focused on the in-situ formation of multiphase gas from the hot ambient phase via thermal instability \citep{Field1965THERMAL1965,Sharma2012ThermalGalaxies,Mccourt2012ThermalHaloes}, and the effects of turbulence \citep{Audit2005,Mohapatra2022CharacterizingSimulations}, magnetic fields \citep{Sharma2010ThermalClusters, Ji2019SimulationsLayers}, metallicity \citep{Das2021ShatterInstability}, density perturbations \citep{Choudhury2019MultiphaseFluctuations}, rotation \citep{Sobacchi2019TheAnalysis}, cosmic rays \citep{Butsky2020TheMedium}, and stratification \citep{Mohapatra2020TurbulenceMedium,Wang2022}. Other studies have examined the evolution of cold gas through mixing and subsequent cooling of the mixed intermediate temperature gas, particularly through cloud-crushing simulations where cold clouds are mixed and accelerated by hot outflow winds \citep[e.g.][]{Armillotta2016,Kanjilal2021GrowthCooling, Abruzzo2022TamingInteractions, Hidalgo-Pineda2023BetterDraping} or cold streams flowing through a hot halo \citep[e.g.][]{mandelkerInstabilitySupersonicCold2020,Ledos2023}.

Across these small-scale studies, it has been found that resolving length scales as small as sub-parsec is crucial for achieving convergence in even the most basic properties, such as the mass distribution across different phases \citep{McCourt2018AGas,Gronke2018TheWind,GronkeTurb2022}. However, accurately simulating these multiphase media in large-scale simulations, which are required to capture larger flows and draw reasonable observational conclusions, remains a significant challenge due to the wide range of scales involved. For example, to properly resolve the observed cold gas structures within a single galactic halo would require a resolution that remains computationally infeasible even for next-generation supercomputers.

There have been many approaches in astrophysical simulations to address this challenge through `subgrid models' that incorporate processes occurring below the resolution limit. Notable examples include models for unresolved turbulence \citep{Schmidt2006, SchmidtFederrath2011A&A...528A.106S, Scannapieco2008ApJ...686..927S}, supernova feedback \citep[e.g.][]{Rosdahl2017MNRAS.466...11R,2016MNRAS.459.2311M}, and star formation \citep[e.g.][]{Federrath2012ApJ...761..156F}. Previous studies, such as \citet{Huang2020} and \citet{Smith2024} (cf. \S~\ref{sec:previous_work} for an overview of previous multiphase subgrid models), have used Eulerian-Lagrangian methods for subgrid treatment of multiphase galactic outflows, where cold gas clouds are treated as particles interacting with the surrounding hot gas. Another possible approach is an Eulerian-Eulerian method, known as the `multifluid' method.

The multifluid approach has been successfully applied to multiphase flows in various terrestrial contexts, including meteorology, combustion processes, and water flows. By tracking multiple fluids on the same grid, these methods allow for explicit representation of interactions between different phases \citep{Prosperetti_Tryggvason_2007}. Recently, \citep{2024MNRAS.535.1672B} implemented a subgrid model with a second pressureless cold fluid for unresolved cold gas clouds. In this study, we explore a different route by using an alternative implementation of the two-fluid method in the astrophysical code \texttt{AREPO} \citep{Springel2010MNRASArepo, Rainer2020AREPOPublic} by \citep{Weinberger2023} to create a subgrid model. Our approach advances the multifluid framework by allowing for arbitrary volume filling fractions of each fluid, the inclusion of physically motivated coupling terms between the phases, and a thorough testing against resolved multiphase simulations.

This paper is structured as follows. In \S~\ref{sec:setup}, we describe the numerical setup used to implement and validate our subgrid model, titled \model. In \S~\ref{sec:model}, we list and explain the ingredients that go into \model before validating it in \S~\ref{sec:verify_kol} and \S~\ref{sec:verify_local} using Kolmogorov scaling and local velocity gradients-based estimates for local turbulent velocities, respectively. We discuss these results in \S~\ref{sec:discussion} before we conclude in \S~\ref{sec:conc}. The visualisations related to this study can be found \href{http://hiteshkishoredas.github.io/research/mogli_subgrid.html}{here}.\footnote{\url{http://hiteshkishoredas.github.io/research/mogli_subgrid.html}}

\section{Numerical Setup}
\label{sec:setup}
We use two simulation setups: the high-resolution resolved single-fluid setup using \texttt{Athena++} \citep{Stone2020} and the analogous \model simulations using multifluid \texttt{AREPO} \citep{Springel2010MNRASArepo, Rainer2020AREPOPublic} framework from \citet{Weinberger2023} with our new subgrid model \model implemented.

\subsection{Turbulent box simulations}

We initialise a cubic computational domain, filled with isobaric hot gas of constant density, $\rho_{\rm hot}$, at a temperature, $T_{\rm hot} = 4\times10^6$K.
Subsequently, we drive turbulence in this box at the largest scale, i.e. the box size ($L_{\rm box}$), using the Ornstein-Uhlenbeck (OU) process \citep{ESWARAN1988257, Schmidt2006} to reach a given turbulent velocity ($v_{\rm turb}$) at steady-state. We let the turbulence driving proceed for about 7$t_{\rm eddy}$, where $t_{\rm eddy} = L_{\rm box}/v_{\rm turb}$ is the eddy-turnover timescale. For the turbulence driving, we set the correlation timescale to $\sim t_{\rm eddy}$, the driving timescale to $0.001 t_{\rm eddy}$ and solenoidal to compressive fraction, $f_{\rm sol} = 0.3$. Note that during the turbulence-driving of the initialisation phase, there is no radiative cooling in \texttt{Athena++} runs, and only a single fluid in \model runs.

At the end of the initialisation phase, we restart the simulation after introducing a dense cold gas cloud with a radius $R_{\rm cloud}$, overdensity $\chi=100$ and temperature $T_{\rm cold} = 4\times10^4$K, in the centre of the box.  The introduction of cloud differs between the single-fluid \texttt{Athena++} and multifluid \model runs, and the details are explained in Sec.~\ref{subsubsec:setup_singlefluid}~\&~\ref{subsubsec:setup_multifluid}. During the turbulence-driving in the initialisation phase, the temperature of the hot gas increases due to turbulence heating. Hence, before introducing the cloud, we rescale the temperature of each cell by a constant factor, to get the average temperature back to $T_{\rm hot}$. We also repeat this two-step process, turbulent driving and the addition of cold cloud, with different random seeds for the turbulence driving to probe the stochasticity of the results.

\subsection{Resolved single-fluid}
\label{subsubsec:setup_singlefluid}

For simulations with \texttt{Athena++} \citep{Stone2020}, we use the default HLLC solver with Piecewise Linear Method (PLM) on primitive variables, RK2 time integrator, adiabatic equation-of-state (EOS) and cartesian geometry. Similar to \citet{Das2024}, we use the CIE cooling curve from \citet{Wiersma2009} at solar metallicity and implement the \citet{Townsend2009} radiative cooling algorithm with a 40-segment power-law fit on the cooling curve for calculating the radiative losses. We stop the cooling and enforce a temperature floor at $T_{\rm floor} = 4\times10^4$K.

We include radiative cooling for one set of simulations (radiative mixing) and do not have radiative cooling for the rest (non-radiative mixing). For introducing the cold gas cloud with an overdensity of $\chi = 100$ in \texttt{Athena++} simulations, we set the density within the cloud region to $\chi \rho_{\rm hot}$ and temperature to $T_{\rm hot}/\chi$, while keeping the local pressure and kinetic energy unchanged. We resolve the dense cold gas cloud by at least 12 grid cells along its diameter, to have converged evolution (\citealp{GronkeTurb2022}; cf. \citealp{Tan2021RadiativeCombustion} for comparison to detailed turbulent mixing layer simulations). We vary the cloud radius ($R_{\rm cloud}$) as well as the $L_{\rm box}/R_{\rm cloud}$ for comparison with analogous multifluid simulations. The dashed circles show the corresponding cold gas cloud size in the simulations.

\subsection{Subgrid multifluid}
\label{subsubsec:setup_multifluid}

\begin{figure*}
    \includegraphics[width=0.9\textwidth]{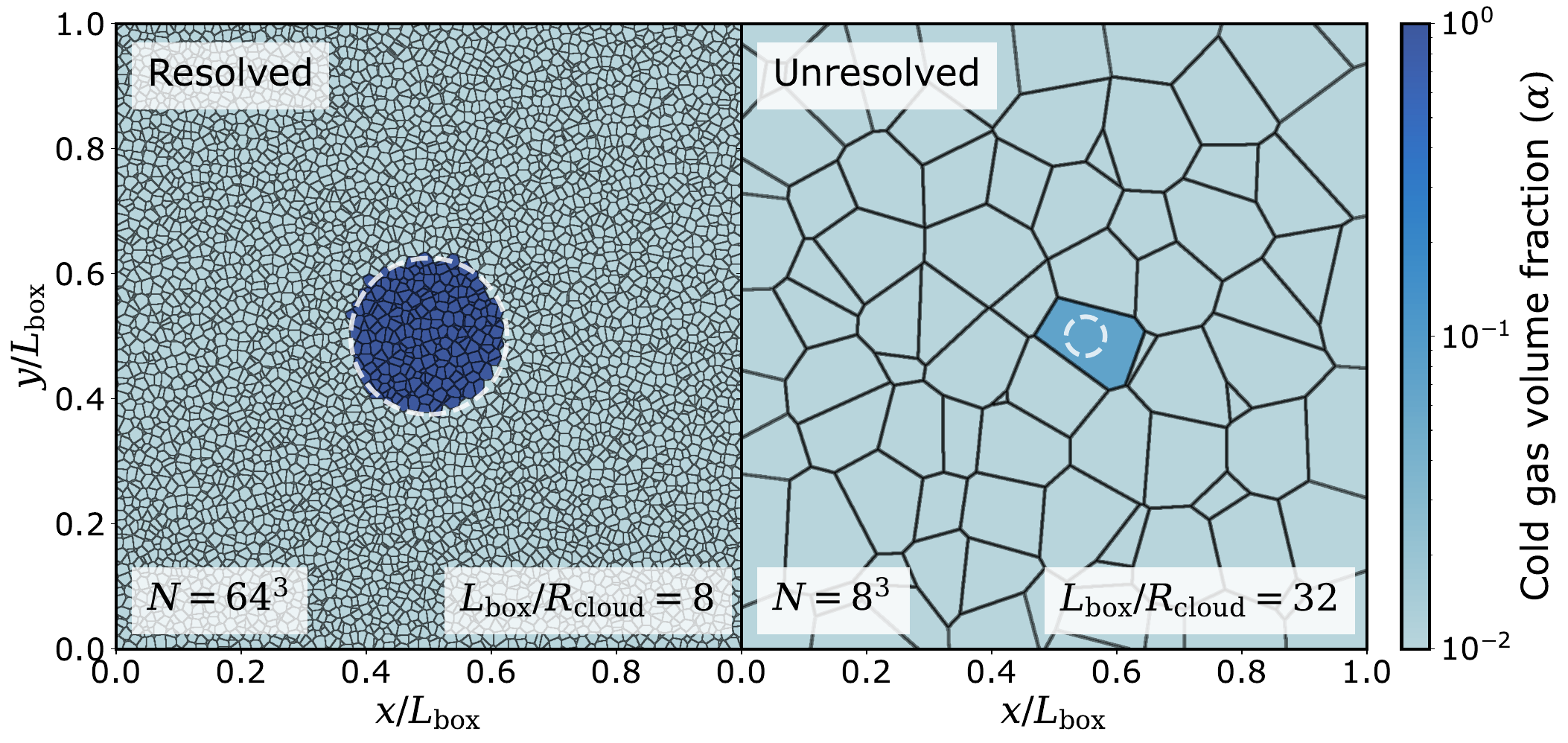}
    \caption{Initial cold fluid volume fraction slices for \model simulations with resolved and unresolved cold gas clouds. The left panel shows an example of a resolved cold gas cloud with $64^3$ cells and $L_{\rm box}/R_{\rm cloud} = 8$, where the cloud is bigger than the grid cells and grid cells inside the volume of the cloud have an $\alpha = 1-\alpha_{\rm floor}$. On the other hand, the right panel shows the initial cold fluid volume fraction for \model simulation with an unresolved cold gas cloud, with $8^3$ cells and $L_{\rm box}/R_{\rm cloud} = 32$. As the cold gas cloud is unresolved, the volume fraction in the cell is set to $\alpha_{\rm floor} + V_{\rm cloud}/V_{\rm cell}$, where $V_{\rm cloud}$ and $V_{\rm cloud}$ are the cloud and grid cell volumes. In both cases, the cells without any cold gas have a volume fraction, $\alpha=\alpha_{\rm floor} = 10^{-8}$. The dashed circles show the corresponding cold gas cloud size in the simulations.}
    \label{fig:IC}
\end{figure*}

For \model simulations with multifluid \texttt{AREPO}, we use the default exact hydrodynamic Riemann solver, time integration, and spatial reconstruction. The details of the multifluid framework are presented in \citet{Weinberger2023}. We use an adiabatic EOS for the hot fluid to allow for turbulent heating and cooling. In contrast, we use a quasi-isothermal EOS for the cold fluid, which resets the internal energy after each timestep, to emulate the temperature floor and fast cooling for cold gas in resolved single-fluid simulations. We use a different (de)refinement criterion where we assume the cell is filled with hot fluid, i.e. we refine depending on $m_{\rm refine} = \sum_{k}m_{k} \rho_{\rm hot}/\rho_{k}$, where $k$ belongs to the set of fluids. We (de)refine the cell to match $m_{\rm refine}$ and a target mass resolution. This (de)refinement criterion ensures that the cells with more cold fluid are not excessively refined and refinement occurs as if we only have the hot fluid.

In a multifluid simulation, we track hot and cold gas as two different fluids. Each cell has a quantity that refers to the fraction of cell volume occupied by the fluids. As we only have two fluids in our setup, we only need to keep track of one volume fraction. We refer to the volume fraction of the cold fluid as $\alpha^\prime$ and that of the hot fluid as $(1-\alpha^\prime)$. We enforce a floor of $\alpha_{\rm floor} = 10^{-8}$ on $\alpha^\prime$ and $(1-\alpha^\prime)$ for numerical reasons. Hence, $\alpha^\prime = \alpha_{\rm floor}$ for a cell filled with hot fluid and $\alpha^\prime = 1-\alpha_{\rm floor}$ for one filled with cold fluid. The $\alpha_{\rm floor}$ dictates the order of magnitude of the smallest amount of cold gas that can be tracked, so it has to be a sufficiently small number. We found that $\alpha_{\rm floor}=10^{-8}$ is small enough, and further decrease does not have a significant effect on the results, apart from the slightly higher computational costs due to stricter source integration tolerances.

After rescaling the temperature (same as the \texttt{Athena++} setup; cf. Sec.~\ref{subsubsec:setup_singlefluid} above), we take the single-fluid output at the end of the turbulence-driving in the initialisation phase and create the multifluid initial condition with the cloud. If the cloud is resolved, we fill the cells within the cloud region with cold fluid by setting $\alpha^\prime = 1 - \alpha_{\rm floor}$. If the cloud is not resolved, we pick the cell closest to the centre of the box and increase the $\alpha^\prime$ by the amount corresponding to the volume of the unresolved cloud. If $V_{\rm cell}$ and $V_{\rm cloud}$ are the volume of the cell and the cloud, respectively, the $\alpha^\prime$ is set to $\alpha_{\rm floor} + V_{\rm cloud}/V_{\rm cell}$. Both fluids have the same velocity at the beginning of the simulations.

Fig.~\ref{fig:IC} shows the initial cold fluid volume fraction slices for \model simulations with resolved and unresolved cold gas clouds. The left panel shows an example of a resolved cold gas cloud with $64^3$ cells and $L_{\rm box}/R_{\rm cloud} = 8$, where the cloud is bigger than the grid cells and grid cells inside the volume of the cloud have an $\alpha = 1-\alpha_{\rm floor}$. On the other hand, the right panel shows the initial cold fluid volume fraction for \model simulation with an unresolved cold gas cloud, with $8^3$ cells and $L_{\rm box}/R_{\rm cloud} = 32$. As the cold gas cloud is unresolved, the volume fraction in the cell is set to $\alpha_{\rm floor} + V_{\rm cloud}/V_{\rm cell}$, where $V_{\rm cloud}$ and $V_{\rm cloud}$ are the cloud and grid cell volumes.

\section{\textbf{\model}: The subgrid model}
\label{sec:model}
Multifluid \texttt{AREPO} evolves multiple fluids on a common grid. This allows for the inclusion of terms for interactions between the fluids. In our case, the interaction terms are mass exchange ($\dot{m}$), momentum exchange ($\dot{\vec{p}}$) and energy exchange ($\dot{E}$) terms. The source terms are integrated using the Bader-Deuflhard semi-implicit integration (\citealp{bader1983semi}; Weinberger et al., in prep). The semi-implicit integrator takes the 10 conservative variables (mass, momenta and energy for each fluid) and integrates the source functions from the subgrid model to calculate the new values over a timestep. \model consists of the model for the source terms for the interactions between the hot and cold gas fluids, which we explain in this section.

\subsection{Definition of the source functions}
\label{subsec:def}

Let $\Q$ and $\Qdot$ denote the conservative variables and source functions, respectively, from the model.
\begin{alignat}{2}
&\Q = \begin{bmatrix}
    m \\
    \vec{p} \\
    E
\end{bmatrix},
&&\Qdot = \begin{bmatrix}
    \dot{m} \\
    \dot{\vec{p}} \\
    \dot{E}
\end{bmatrix}
\end{alignat}

We split the source functions into three components, that refer to contributions from three different physical processes. The first contribution, $\Qdrag$, is due to the hydrodynamic drag between fluids. The drag interaction does not lead to a mass exchange but can result in momenta and energy exchange. The second one, $\Qmix$, is from the mixing between cold into hot fluid, and the third, $\Qgrow$, is from the cooling of the mixed gas from hot fluid to cold fluid. Both second- and third-source function contributions involve mass, momenta, and energy exchange. So, the full source function can be written as
\begin{equation}
    \Qdot = \Qdrag + \Qmix + \Qgrow.
    \label{eq:general_source_terms}
\end{equation}

While mass and momentum are conserved for all the different parts of the source functions,  energy is only conserved for $\Qdrag$ and $\Qmix$. For $\Qgrow$, the energy is not conserved as thermal energy is dissipated via radiative cooling and the difference between the energy exchange is given by $\dot{E}_{\rm cooling}$. We can write the conservation relations as
\begin{alignat}{2}
    & \Qdrag[{\rm cold}] &&= -\Qdrag[{\rm hot}], \\
    & \Qmix[{\rm cold}] &&= -\Qmix[{\rm hot}], \\
    & \Qgrow[{\rm cold}] &&= -\Qgrow[{\rm hot}] - \begin{bmatrix}
    0 \\
    0 \\
    \dot{E}_{\rm cooling}
\end{bmatrix} .
\end{alignat}

From here onwards, we will denote elements of the `cold' or `hot' source functions with ${}_{\rm cold}$ or ${}_{\rm hot}$, respectively.

\subsection{The cold gas volume filling fraction}
As volume fraction is not part of the 10 conservative variables, we need to calculate the volume fraction from the conservative quantities.
\begin{alignat}{2}
    &\alpha^\prime &&= \frac{m_{\rm cold} u_{\rm cold}}{m_{\rm cold} u_{\rm cold} + m_{\rm hot} u_{\rm hot}}
\end{alignat}
where $u_{\rm hot}$ and $u_{\rm cold}$ are the specific internal energy of the hot and cold fluid, respectively. As $\alpha$ has a floor and ceiling value, we redefine the volume fraction such that values between $\alpha_{\rm floor}$ and $\alpha_{\rm ceil}$ map linearly to their physical values in $[0, 1]$. So, a cell-filled with hot fluid and $\alpha^\prime = \alpha_{\rm floor} $ corresponds to $\alpha = 0$, while a cell filled with cold fluid and $\alpha^\prime = 1-\alpha_{\rm floor} $ corresponds to $\alpha = 1$. This new mapped value is subsequently used for calculations in the source functions.
\begin{alignat}{2}
    &\alpha &&= \begin{cases}
        0.0 &\text{for } \alpha^\prime < \alpha_{\rm floor} \\
        \dfrac{\alpha - \alpha_{\rm floor}}{1- 2\alpha_{\rm floor}} &\text{for } \alpha \in [\alpha_{\rm floor}, 1-\alpha_{\rm floor}] \\
        1.0 &\text{for } \alpha^\prime > 1-\alpha_{\rm floor}
    \end{cases}
\end{alignat}

\subsection{Drag forces}
\label{subsec:drag}

One obvious change of momentum and energy stems from the hydrodynamic drag between the fluids in a cell, in the presence of a relative velocity $\Delta \vec{v}$ giving rise to $\dot Q_{\rm drag}$ in Eq.~\eqref{eq:general_source_terms}. The drag force between the two fluids in the cell will be
\begin{alignat}{2}
    \dot p_{\rm drag,cold} = &-\vec{F}_{\rm drag} &&=  0.5~C_{\rm D}~\rho_{\rm hot}~A_{\rm cross}(\alpha)~|\Delta \vec{v}| \Delta \vec{v}
\end{alignat}
where $C_{\rm D} = 0.5$ is the drag coefficient and $A_{\rm cross}(\alpha)$ is the cross-sectional area of the cold fluid in the direction of the relative velocity. We further discuss the functional form of $A_{\rm cross}$ (Eq.~\eqref{eq:A_cross}) in Section~\ref{subsec:area_factors}.

Drag force also leads to an exchange of energy between the two fluids. The rate of energy exchange is set to,
\begin{equation}
\dot E_{\rm cold} = -\vec{F}_{\rm drag}.\vec{v}_i.
\end{equation}
Here, $\vec{v}_i$ is the centre of mass velocity, which will be attained at equilibrium \citep{SaurelAgrall1999}, i.e.,
\begin{alignat}{2}
    &\vec{v}_{\rm i} &&= \frac{m_{\rm cold} \vec{v}_{\rm cold} + m_{\rm hot} \vec{v}_{\rm hot}}{m_{\rm cold} + m_{\rm hot}} = \frac{\vec{p}_{\rm cold} + \vec{p}_{\rm hot}}{m_{\rm cold} + m_{\rm hot}}.
\end{alignat}

As mentioned earlier, the drag force contribution does not cause mass exchange. Hence, the final form of $\dot{\mathbf{Q}}_{\rm drag, cold}$ is
\begin{alignat}{2}
    & \Qdrag[{\rm cold}] &&= \begin{bmatrix}
        0 \\
        -\vec{F}_{\rm drag} \\
        -\vec{F}_{\rm drag}.\vec{v}_{\rm i}
    \end{bmatrix}.
\end{alignat}
Later, in Sec.~\ref{subsec:ver_adiab_mix}~\&~\ref{sec:verify_local} we include $\Qdrag$ and verify the \model model.

\begin{figure}
    \centering
    \includegraphics[width=\linewidth]{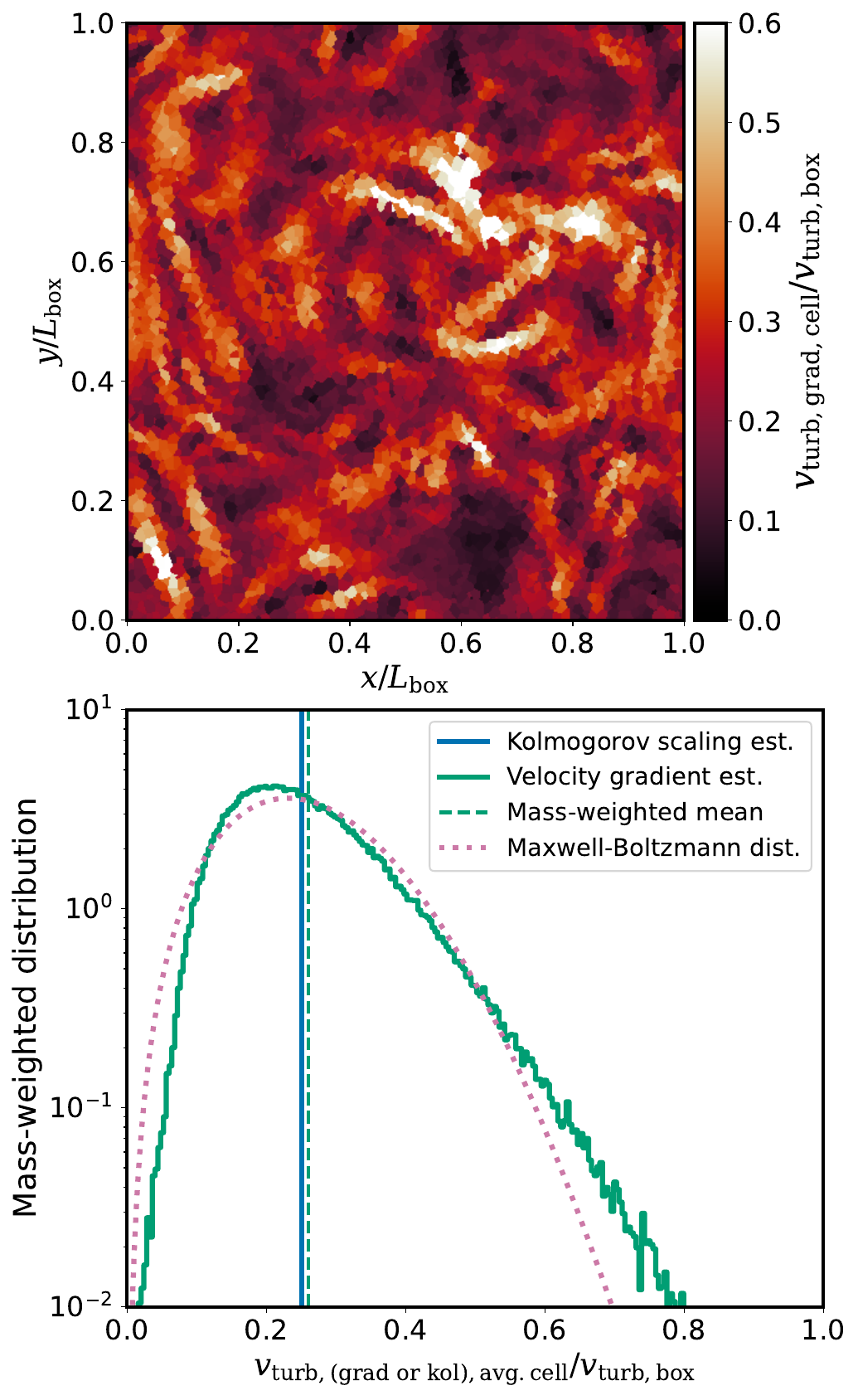}
    \caption{\textit{Top panel} shows a slice of $v_{\rm turb, grad}$ from a simulation with a turbulent Mach number, $\mathcal{M}_{\rm turb, box} = 0.5 $ at the box scale. It shows how the velocity gradient-based estimation (\texttt{grad}) can capture the spatial variation in the local velocity dispersion, in other words, the local turbulent velocity. \textit{Bottom panel} shows, in solid lines, the distribution of the local turbulent velocity, at the scales of average cell size instead of local cell size, in the same snapshot as the top panel. We find that while the mass-weighted mean of from \texttt{grad} method, shown as the dashed green line, agrees with the estimate from \texttt{kol}, shown as the solid blue line. The pink dotted line shows the expected Maxwell-Boltzmann distribution with the same mean as the mass-weighted mean from \texttt{kol} method. Even though the mean turbulent velocity from the two simulations are very similar, the distribution of velocities is drastically different, with the \texttt{kol} method leading to a fixed value for a fixed length scale, and the \texttt{grad} method matching the expected Maxwell-Boltzmann distribution.}
    \label{fig:kol_vs_grad}
\end{figure}

\subsection{Turbulent Mixing}
\label{subsec:mixing}

Next, we consider the contribution to source functions from the turbulent mixing of cold gas into hot gas. This part of the source function contains mass, momenta and energy exchange and is captured by $\mathbf{Q}_{\rm mix}$
 (cf. Eq.~\eqref{eq:general_source_terms}).

To first order, we expect the turbulent mixing to destroy the cold gas on a characteristic timescale $t_{\rm destroy}$, i.e., $\dot m_{\rm cold\rightarrow hot}\sim m / t_{\rm destroy}$.
From previous work as well as analytical considerations, we expect this destruction timescale to be the Kelvin-Helmholtz or Rayleigh-Taylor timescale of the cloud \citep[e.g.][]{Klein1994,GronkeTurb2022}
\begin{equation}
    t_{\rm destroy} = \dfrac{\chi^{1/2}l_{\rm cold}}{v_{\rm turb}}.
    \label{eq:t_destroy}
\end{equation}
Here, $l_{\rm cold}$ is the effective size of the cold gas in the cell
\begin{equation}
    l_{\rm cold} = \bigg(\frac{\alpha V_{\rm cell}}{4\pi/3}\bigg)^{1/3},
\end{equation}
$v_{\rm turb}$ is the turbulent velocity, and $\chi$ overdensity of a given cell
\begin{equation}
    \chi \equiv \frac{\rho_{\rm cold}}{\rho_{\rm hot}} = \bigg(\frac{1}{\alpha} -1\bigg)\frac{m_{\rm cold}}{m_{\rm hot}}.
\end{equation}
Note that while $\chi$ is independent of $\alpha$ for a given cell, $l_{\rm cold}$ strongly depends on it and thus gives rise to a short destruction time for, e.g., a predominantly hot cell. We explicitly verified this destruction term in Sec.~\ref{subsec:ver_adiab_mix}~\&~\ref{sec:verify_local}. While we calculate $l_{\rm cold}$ as the size of a monolithic cold cloud in the cell, it refers to the length-scale of the cold gas which we subsequently use to calculate destruction timescale, $t_{\rm destroy}$ in Eq.~\eqref{eq:t_destroy}, it does not relate to the cold gas structure within the cell.

Within any cell, we expect the mass exchange to slow down and drop with increasing volume fraction of the cold fluid, due to the decrease in the interface area between the two fluids, over which the exchange can happen; for instance, we expect a cold gas cell surrounded by other gas cells with a $\sim 1$ cold gas volume fraction this mass exchange rate to be $\sim 0$ as the cell is entirely `shielded' from the hot medium. We include an extra multiplicative factor, i.e. the area factor, of $2h(\alpha)$ to account for this dependence on the interface area. This factor essentially encodes the details of the subgrid cold gas structure. The exact form of this factor is derived and discussed in \S~\ref{subsec:area_factors}\footnote{Note that we deliberately did not absorb the factor $2$ into the fudge factor to facilitate comparison with prior work; cf. \S~\ref{subsec:area_factors}}.

\begin{figure}
    \centering
    \includegraphics[width=\linewidth]{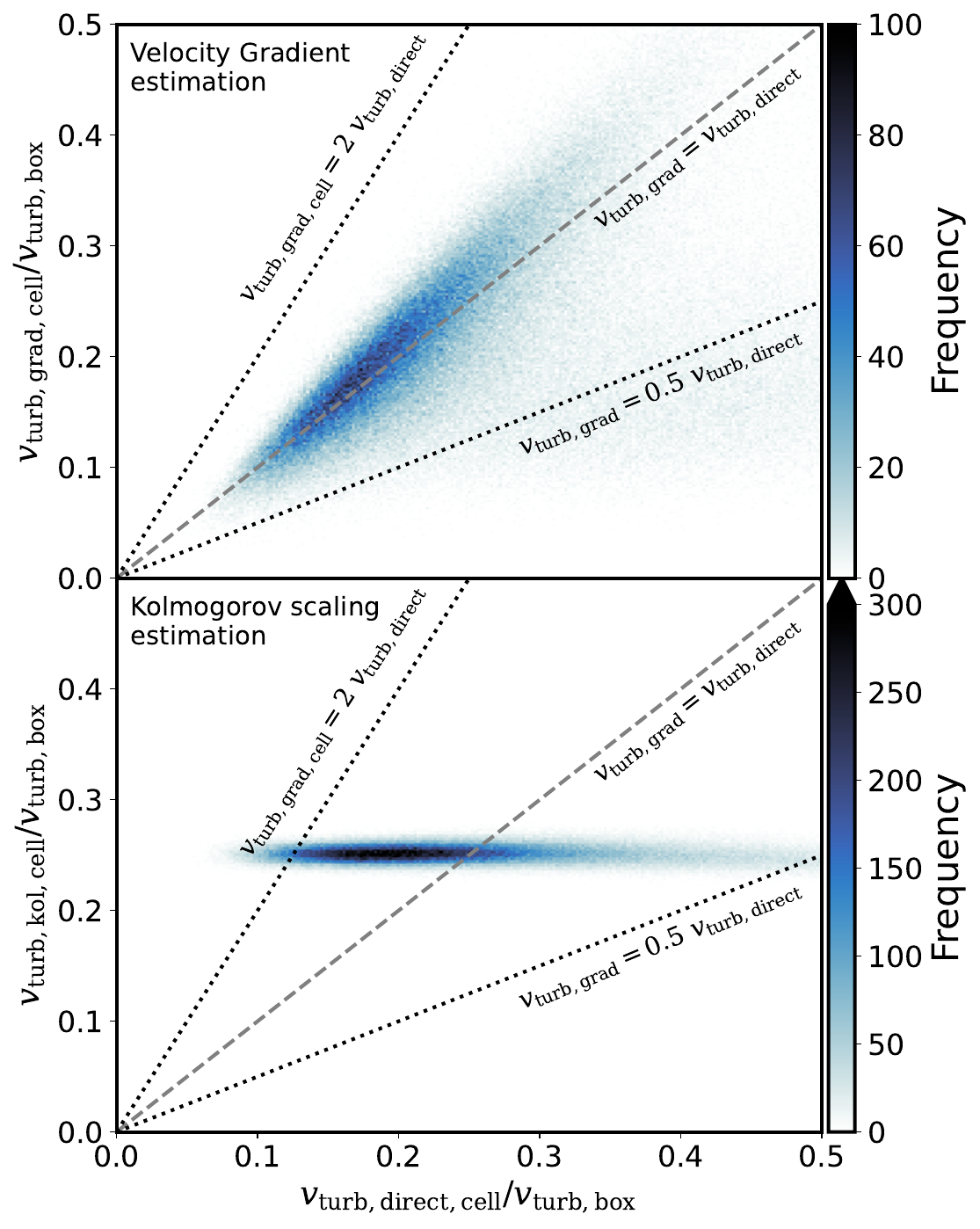}
    \caption{The comparison between the directly calculated velocity dispersion ($v_{\rm turb, direct}$) and the approximated local velocity dispersions using both estimation methods. The top panel shows the comparison with the velocity gradient-based method (\texttt{grad}) and the bottom panel show the comparison with the Kolmogorov spectrum-based method (\texttt{kol}).}
    \label{fig:arepo_test_kolgrad}
\end{figure}

We find that solely the area factor, $2h(\alpha)$), is inadequate to suppress the cold gas destruction in cells which have a high cold fluid mass fraction, $\alpha_{\rm mass} = m_{\rm cold}/(m_{\rm cold} + m_{\rm hot})$, especially for cases with resolved cold gas.
This can be easily understood through additional `shielding' as those predominantly cold gas cells are mostly also surrounded by other predominantly cold gas cells preventing their destruction.
The effect is that the mass exchange from cold to hot occurs only in cells that possess a mass fraction $\alpha_{\rm mass}$ less than a threshold. We find that a $\alpha_{\rm mass}$ threshold of $0.15$ works well across all tests (\S~\ref{subsec:ver_adiab_mix}), i.e.,
\begin{alignat}{2}
    &\dot{m}_{\rm cold\rightarrow hot} &&= \begin{cases}
    2h(\alpha) \dfrac{m_{\rm cold}}{t_{\rm destroy}}  & \alpha_{\rm mass} < 0.15 \\
    0  & \text{otherwise}. \\
    \end{cases}
    \label{eq:mdot_mix}
\end{alignat}

The momenta and energy exchange are the corresponding fluxes caused by the mass exchange. Hence, we get a source function contribution due to mixing,
\begin{alignat}{2}
    & \Qmix[{\rm cold}] &&= \begin{bmatrix}
        -\dot{m}_{\rm cold\rightarrow hot} \\
        - \vec{v}_{\rm cold} \dot{m}_{\rm cold\rightarrow hot} \\
        -\frac{1}{2} \dot{m}_{\rm cold\rightarrow hot} v_{\rm cold}^2 - \dot{m}_{\rm cold\rightarrow hot} u_{\rm cold}
    \end{bmatrix}
\end{alignat}

\subsection{Cold gas growth}
\label{subsec:grow}

Finally, we include the source function contribution due to the cooling of mixed gas to create more cold gas. Previous studies like \citet{GronkeTurb2022} have shown that for cases that are well within the growth regime, the cold gas mass follows an exponential growth, with a characteristic timescale of $t_{\rm grow}\sim \chi (t_{\rm cool} l/v_{\rm turb})^{1/2}$, i.e., characterised by the geometric mean of the cold gas mixing and cooling time, and where $l$ refers to the cold gas structure lengthscale and $v_{\rm turb}$ is the turbulent velocity at the lengthscale $l$. Note that this growth time is backed up by analytical arguments of combustion theory and numerically verified using small-scale simulations \citep{Tan2021RadiativeCombustion}. We can rewrite the expression for $t_{\rm grow}$ in terms of the local cell properties as,
\begin{alignat}{2}
    &t_{\rm grow} &&= \chi (t_{\rm destroy} t_{\rm cool, cold})^{1/2} \alpha^{1/9}
    \label{eq:tgrow_implemented}
\end{alignat}
where the additional factor $\alpha^{1/9}$ arises to account for the effective cold gas size within a cell.

Both $t_{\rm destroy}$ and $t_{\rm cool, cold}$ can be calculated from local grid cell properties, but in our tests, we consider a fixed value of $t_{\rm cool, cold}$. We use the value of $t_{\rm cc}$ and the required ratio if $t_{\rm cool, cold}/t_{\rm cc}$ to obtain the required value of $t_{\rm cool, cold}$.
We also include the area factor of $2h(\alpha)$ for this interaction, same as Eq.~\eqref{eq:mdot_mix}, which gives the hot to cold fluid mass exchange rate,
\begin{alignat}{2}
    &\dot{m}_{\rm hot\rightarrow cold} &&= 2h(\alpha) \frac{m_{\rm cold}}{t_{\rm grow}}
    \label{eq:mdot_grow}
\end{alignat}
As mentioned earlier, the energy across the system is not conserved during cold gas growth. The difference in internal energy during the mass exchange from hot to cold fluid is radiated away and hence subtracted
\begin{alignat}{2}
    & \dot{E}_{\rm cooling} &&= \dot{m}_{\rm hot\rightarrow cold} (u_{\rm hot}-u_{\rm cold}).
\end{alignat}

Including the momentum exchange due to mass exchange, we obtain the expression for the source function contribution due to cold gas growth
\begin{alignat}{2}
    & \Qgrow[{\rm cold}] &&= \begin{bmatrix}
        \dot{m}_{\rm hot\rightarrow cold} \\
        \vec{v}_{\rm hot} \dot{m}_{\rm hot\rightarrow cold} \\
        \frac{1}{2} \dot{m}_{\rm hot\rightarrow cold} v_{\rm hot}^2 + \dot{m}_{\rm hot\rightarrow cold} u_{\rm hot}
    \end{bmatrix} - \begin{bmatrix}
    0 \\
    0 \\
    \dot{E}_{\rm cooling}
    \end{bmatrix}.
\end{alignat}

\subsection{Turbulent velocity estimation}
\label{subsec:vturb}

Turbulence is key in mixing hot and cold material, and thus, the expressions in $\Qmix$ and $\Qgrow$ depend on the local turbulent velocity at the scale of cold fluid in the cell, as $v_{\rm turb} = v_{\rm turb, cell} (l_{\rm cold}/l_{\rm cell})^{1/3}$, where $l_{\rm cell}$ is the cell size. The most accurate way to accomplish this is to have a turbulence model that keeps track of the subgrid turbulence (akin to \citealp{Schmidt2006} or \citealp{Semenov2024}). As the implementation of such a model in a moving-mesh code like \texttt{AREPO} is out of the scope of this study, we use two methods to approximate the turbulence at the grid cell scales.

\begin{figure}
    \includegraphics[width=\columnwidth]{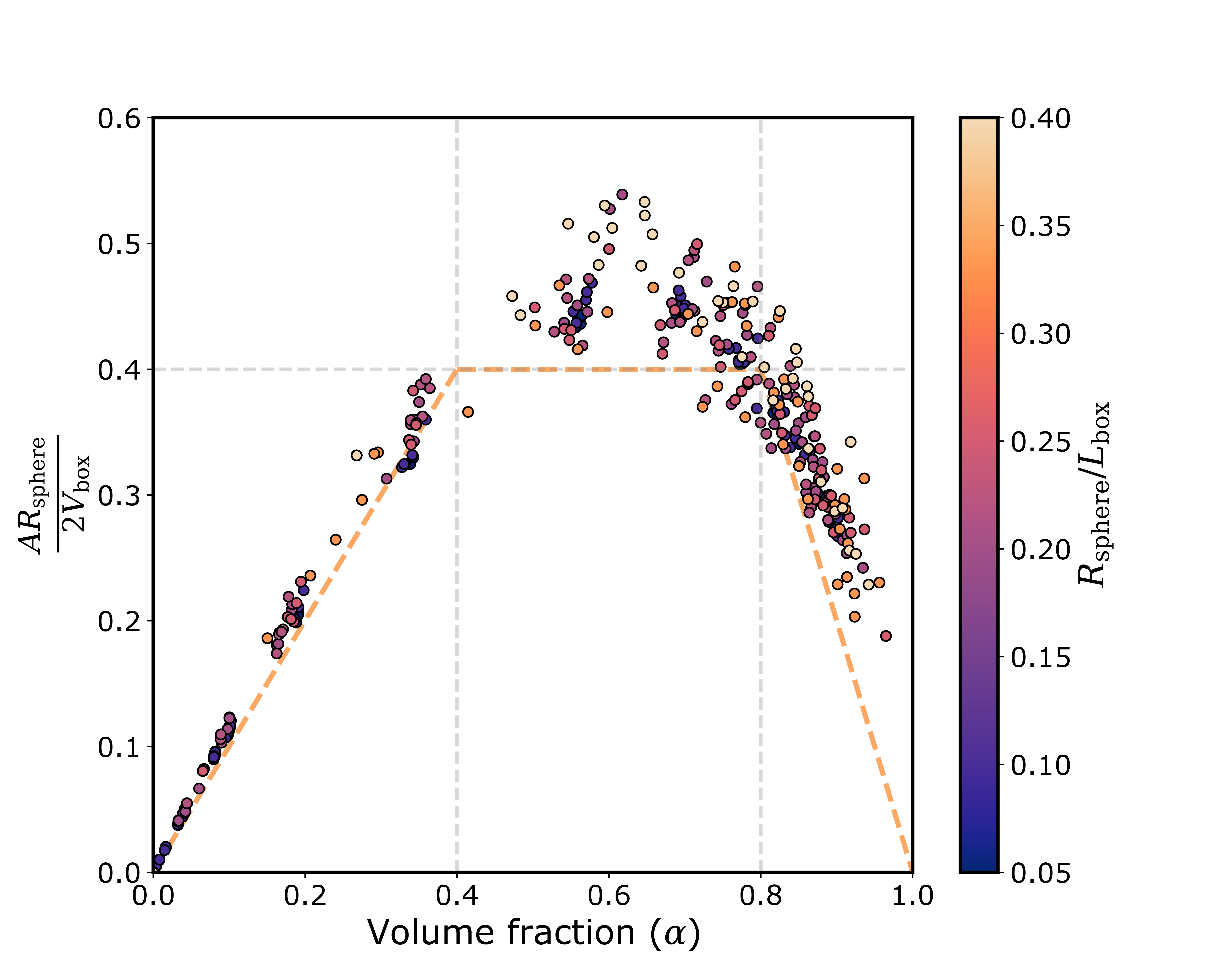}
    \caption{Variation of $AR/(2V_{\rm box})$ with volume fraction ($\alpha$) in a 3D box. The colour of the points shows the size of the individual spheres, relative to the box size and the orange lines correspond to the approximate fit for the points (Eq.~\eqref{eq:interface_area}).}
    \label{fig:ar_vs_volfrac}
\end{figure}

\subsubsection{Estimation by Kolmogorov Scaling (\texttt{kol})}
The first method (\texttt{kol}) assumes that the turbulence is fully developed and follows the Kolmogorov spectrum to the subgrid scales. So, by scaling the box-scale turbulent velocity ($v_{\rm turb, box}$) down to the grid cell scale, we can estimate the local turbulence.
\begin{alignat}{2}
    &v_{\rm turb, \texttt{kol}, cell} &&= v_{\rm turb, box} \bigg(\frac{l_{\rm cell}}{L_{\rm box}}\bigg)^{1/3}
    \label{eq:turb_kol}
\end{alignat}
This method is feasible for setups with fully developed turbulence at a resolved scale, like our turbulent box simulations. However, it is not appropriate in simulations with significant spatial and/or temporal variation in turbulence.

\subsubsection{Estimation by velocity gradients (\texttt{grad})}
\label{subsubsec:grad_vturb}
The second method of local turbulence estimation (\texttt{grad}) avoids the problems of the \texttt{kol} method by estimating the local velocity dispersion at any given time. We use the Jacobian of the velocity vector $(\mathbf{J}_{ij} =  \partial v_i/\partial x_j)$, to estimate velocity dispersion at the grid scale $(\sigma_{\vec{v}, \rm cell})$, and assume a fully-developed Kolmogorov spectrum on smaller scales to estimate the turbulent velocity at the subgrid cold gas scale ($l_{\rm cold}$) from the calculated grid-scale.

As explained in \citet{Pakmor2016}, in \texttt{AREPO}, the gradient of a quantity, say $\phi$, in the $i^{\rm th}$ cell ($\vec{\nabla}_i\phi$) is calculated to be the best fit for Eq.~\eqref{eq:grad}, over all neighbouring cells. Let the $i^{\rm th}$ have $n_{\rm ngb}$ neighbouring cells, each denoted by a subscript $j$,
\begin{alignat}{2}
&\phi_j &&= \phi_i + \vec{\nabla}_i \phi.\vec{x}_{ji} \label{eq:grad}.
\end{alignat}
We use Eq.~\eqref{eq:grad} to calculate the mean of $\phi$ over the neighbouring cells in $n_{\rm ngb}$ including the $i^{\rm th}$ cell,
\begin{alignat}{2}
&\mu_i(\phi_j) &&= \phi_i  +\vec{\nabla}_i\phi.\mu_i(\vec{x}_{ji}). \label{eq:mean}
\end{alignat}
Subsequently, we use the Eq.~\eqref{eq:mean} for the mean to calculate the standard deviation for $\phi$, $\sigma_{\phi}^2$.
\begin{alignat}{2}
    &\sigma_{\phi}^2 &&= \frac{1}{n_{\rm ngb}+1} \sum^{n_{\rm ngb}}_j\bigg[\vec{\nabla}_i \phi.\left(\vec{x}_{ji} - \mu_i(\vec{x}_{ji})\right)\bigg]^2
   \label{eq:std}
\end{alignat}

\begin{figure}
    \includegraphics[width=\columnwidth]{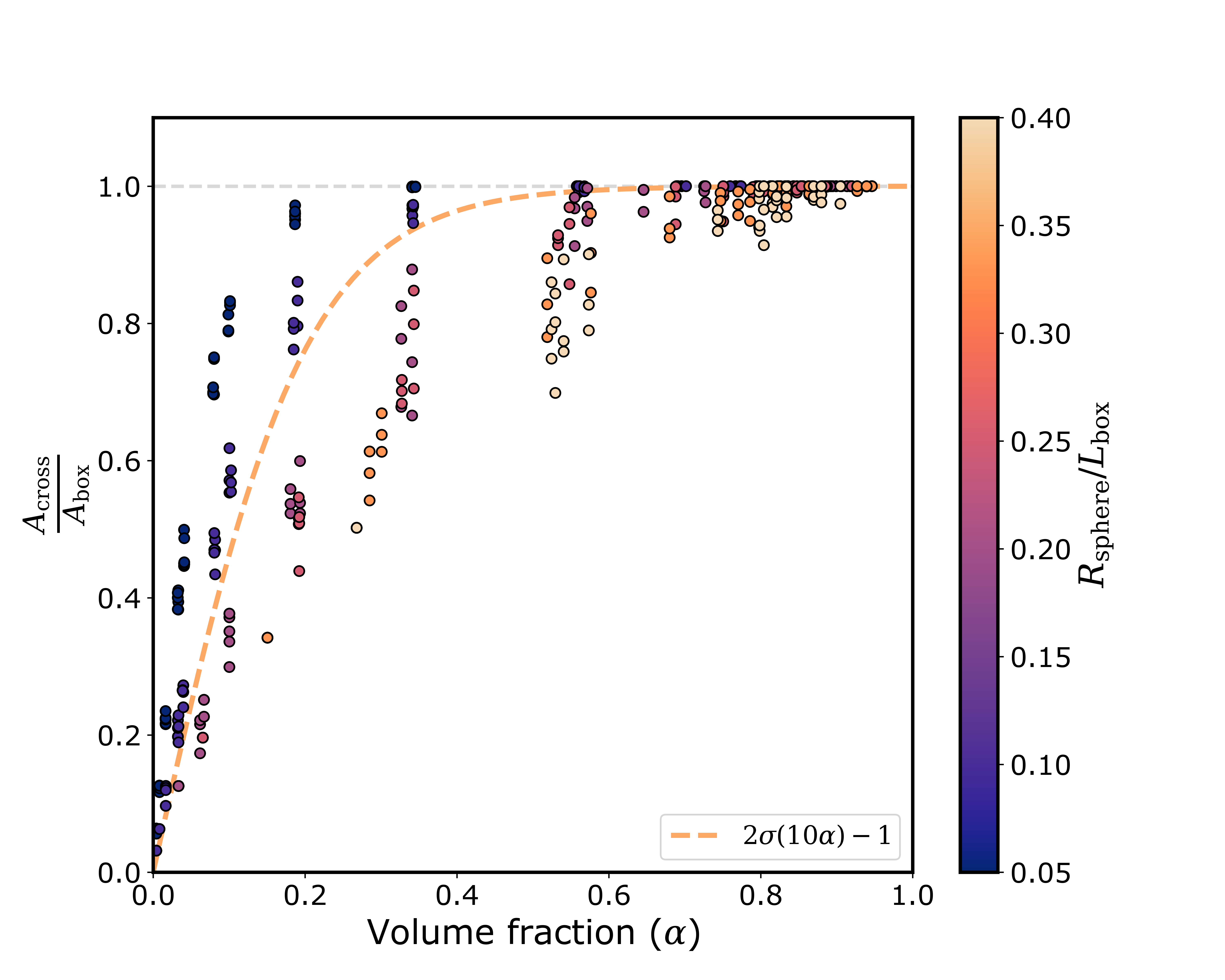}
    \caption{Variation of $A/A_{\rm box}$ with volume fraction ($\alpha$) in a 3D box. The colour of the points shows the size of the individual spheres, relative to the box size and the orange lines correspond to the approximate fit for the points (Eq.~\eqref{eq:A_cross}).}
    \label{fig:cross_section_area}
\end{figure}
To evaluate the Eq.~\eqref{eq:std} further, we need to make some assumptions.
First, we assume that the neighbouring cells are uniformly distributed around the current cell. Hence, the weighted average of the displacements, $\mu_i(\vec{x}_{ji})$ is expected to have a marginal magnitude and can be ignored. Next, we assume that the neighbouring cells are at roughly equal distances of $x_{\rm ngb}$ from the current cell, and we can replace $\vec{x}_{ji}$ with $x_{\rm ngb} \hat{x}_{ji}$. Here, $\hat{x}_{ji}$ refers to the unit vector in the direction of $\vec{x}_{ji}$, i.e. $\hat{x}_{ji} = \vec{x}_{ji}/|x_{ji}|$. This simplifies Eq.~\eqref{eq:std} to
\begin{alignat}{2}
    &\sigma_{\phi}^2 &&= \frac{x_{\rm ngb}^2}{n_{\rm ngb}+1} \sum^{n_{\rm ngb}}_j\bigg[\vec{\nabla}_i \phi.\hat{x}_{ji}\bigg]^2.  \label{eq:vel_disp_discrete}
\end{alignat}
Eq.~\eqref{eq:vel_disp_discrete} shows that the velocity dispersion is related to the root-mean-square value of the component of gradient towards the neighbouring cells. The exact value of this quantity will depend on the relative position and number of the neighbouring cells. Hence, to further simplify the Eq.~\eqref{eq:vel_disp_discrete}, we postulate that this value is close to the value at the limit of $n_{\rm ngb} \rightarrow \infty$, with all the neighbouring cells uniformly distributed around the current cell. This allows us to rewrite Eq.~\eqref{eq:vel_disp_discrete} in its integral form and to obtain the value at the limit by evaluating the integral,
\begin{align}
     \sigma_{\phi, n_{\rm ngb} \rightarrow \infty}
    = \sqrt{\frac{1}{3}}~x_{\rm ngb}|\nabla_i\phi|
    \label{eq:integral_std}
\end{align}
To replace $\phi$ with $v_x$,$v_y$ and $v_z$ \footnote{Note that the velocity gradients used in the \model simulations are the slope-limited gradients that are used in the finite-volume solver.} in Eq.~\eqref{eq:integral_std}, we use the approximation $x_{\rm ngb}\approx V_{\rm cell}^{1/3}$. This yields an estimate of the local turbulence
\begin{alignat}{1}
    &v_{\rm turb, \texttt{grad}, cell} = (\sigma_{v_x}^2 +\sigma_{v_y}^2 +\sigma_{v_z}^2)^{1/2}
    = V_{\rm cell}^{1/3}\sqrt{\frac{1}{\xi}\sum_{i, j}^3\left(\frac{\partial v_j}{\partial x_i}\right)^2} \label{eq:final}
\end{alignat}
where, $\xi = 3$.
For multifluid simulations, if there are $n_k$ fluids, we evaluate the local turbulent velocity with Eq.~\eqref{eq:final} for each fluid separately and take the mass-weighted average as the final turbulent velocity estimate per cell, i.e.,
\begin{alignat}{1}
    &v_{\rm turb, \texttt{grad}, cell} = \dfrac{1}{m_{\rm cell}}\sum_k^{n_k} m_k v_{\rm turb, \texttt{grad}, k}
    \label{eq:final_multifluid}.
\end{alignat}

\begin{figure}
    \includegraphics[width=\columnwidth]{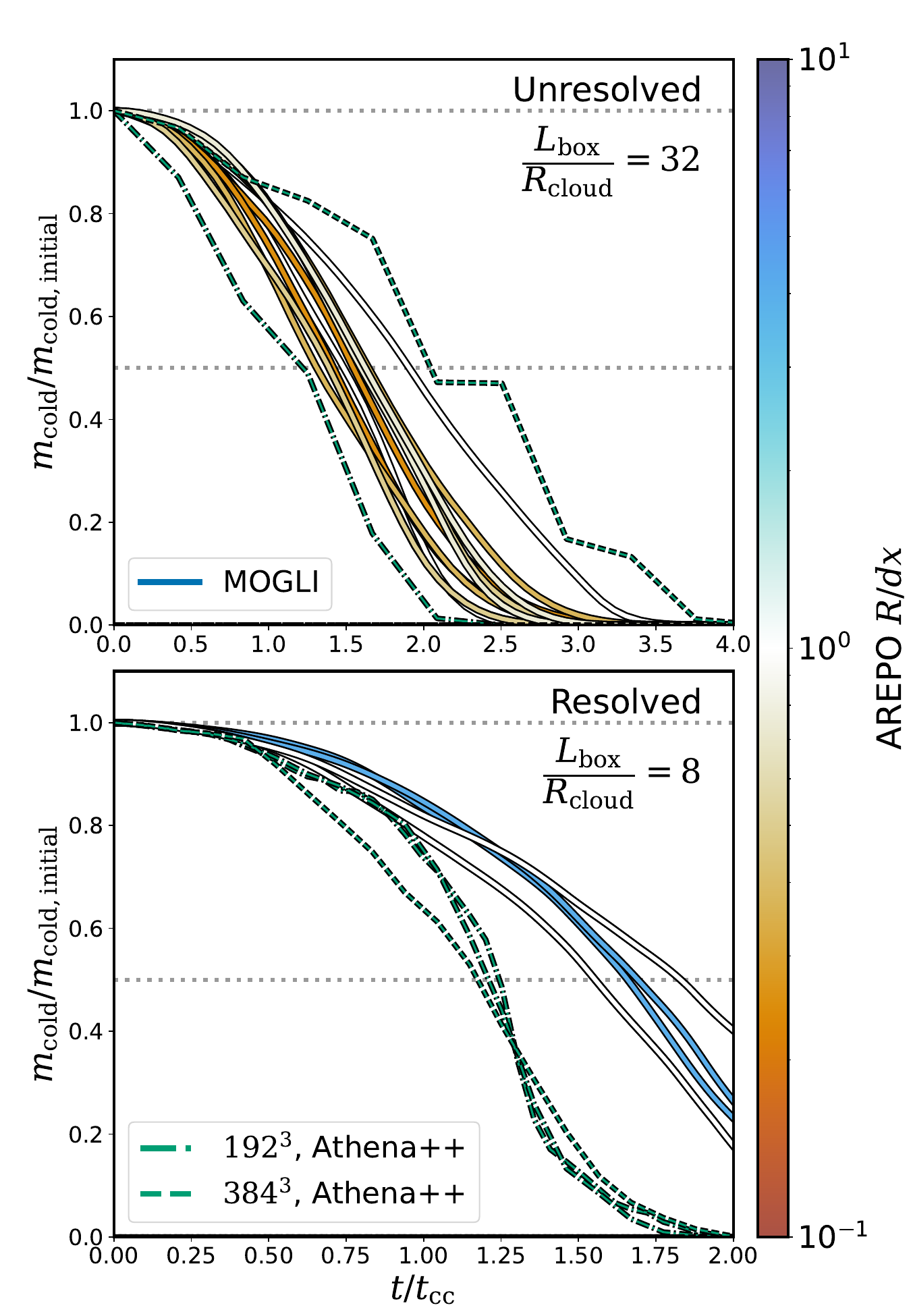}
    \caption{Cold gas evolution in non-radiative MOGLI runs with time, normalised to the initial cloud-crushing time ($t_{\rm cc}$), with $\mathcal{M}_{\rm turb} = 0.5$. The solid lines show the cold gas evolution, as the total mass of the cold fluid, with the colour of the line denoting the initial $R_{\rm cloud}/dx$. The dot-dashed and dashed lines show the cold gas evolution in the benchmark \texttt{Athena++} simulations, with resolutions $192^3$ and $384^3$ respectively. \textit{Top panel} shows the evolution for simulations with unresolved initial cloud $L_{\rm box}/R_{\rm cloud} = 32$ and \textit{bottom panel} shows the same for resolved initial cloud $L_{\rm box}/R_{\rm cloud} = 8$. This shows that the cloud destruction timescales in \model are in agreement with the timescales in benchmark \texttt{Athena++}.}
    \label{fig:cold_gas_adiab}
\end{figure}

\subsubsection{Comparison and validation of the methods}
The top panel of Fig.~\ref{fig:kol_vs_grad} shows a slice of $v_{\rm turb, grad, cell}$ from a simulation with a turbulent Mach number, $\mathcal{M}_{\rm turb, box} = 0.5 $ at the box scale using $N_{\rm cells}=64^3$ resolution elements. It shows how the velocity gradient-based estimation (\texttt{grad}) can capture the spatial variation in the local velocity dispersion, in other words, the local turbulent velocity. The bottom panel shows, in solid lines, the estimation for local turbulent velocity at the lengthscale of the average cell size ($l_{\rm cell, avg} = (V_{\rm box}/N_{\rm cell})^{1/3}$) in the same snapshot as the top panel. As the \texttt{kol} only depends on the lengthscale, it returns a fixed local turbulent velocity for a fixed lengthscale. The blue solid line shows the \texttt{kol} method estimate for the $l_{\rm cell, avg}$. The green solid line shows the distribution of the \texttt{grad} estimates for local turbulent velocities at scale of $l_{\rm cell}$, and the green dashed line shows the corresponding mass-weighted mean. We find that the estimate from \texttt{kol} and mass-weighted mean from \texttt{grad} agree very well with each other, and the expected value $\bar v_{\rm turb} \approx v_{\rm turb,box} / N_{\rm cells}^{1/3}$. 

As the turbulence velocities along each basis directions ($v_{\mathrm{turb}, x}$, $v_{\mathrm{turb}, y}$, $v_{\mathrm{turb}, z}$) roughly follow gaussian distribution, the turbulent velocity magnitude $v_{\rm turb} = (v_{\mathrm{turb}, y}^2 + v_{\mathrm{turb}, y}^2 + v_{\mathrm{turb}, z}^2)^{1/2}$ is expected to follow a Maxwell-Boltzmann distribution. In Fig.~\ref{fig:kol_vs_grad}, the pink dotted line shows the expected Maxwell-Boltzmann distribution, with the same mean as the the mass-weighted mean from \texttt{grad} method. We find that the distribution of local turbulent velocity estimate from \texttt{grad} matches well with the Maxwell-Boltzmann distribution, with some deviations at small and large velocity magnitudes. The deviations at small velocity magnitudes are likely due to the resolution limit and lack of small scales, while the deviations at the larger velocities are probably due to slope-limiting for the gradients.

In order to test the ability of Eq.~\eqref{eq:final_multifluid} to approximate the local velocity dispersion, we calculate the velocity dispersion at grid cell size in the neighbourhood for each grid cell in the simulation, $v_{\rm turb, direct}$ and use it as the benchmark we want to approximate. As the directly calculated velocity dispersion is over the whole neighbourhood with a volume $V_{\rm ngb} = \sum_{\rm ngb} V_{\rm cell, j}$, we use the Kolmogorov scaling to scale it down to the grid cell size as $v_{\rm turb, direct} = \sigma_{\rm turb, ngb} ( V_{\rm cell}/V_{\rm ngb})^{1/9}$ where the latter term introduces corrections of order unity.

Fig.~\ref{fig:arepo_test_kolgrad} shows the comparison between the directly calculated velocity dispersion ($v_{\rm turb, direct}$) and the approximated local velocity dispersions using both estimation methods. The top panel shows the comparison with the velocity gradient-based method (\texttt{grad}) and the bottom panel show the comparison with the Kolmogorov spectrum-based method (\texttt{kol}).
For the bulk of the cells, $v_{\rm turb, direct}$ and $v_{\rm turb, grad}$ agree well with each other and are within a factor of 2 between each other. The slight deviations from $v_{\rm turb, direct}$ are likely due to the assumptions involved in obtaining the Eq.~\eqref{eq:final}. On the other hand, the $v_{\rm turb, kol}$ from the \texttt{kol} method is unable to capture the spatial variations in the $v_{\rm turb, direct}$.

During our non-radiative turbulent mixing tests, explained later in Sec.~\ref{sec:verify_local}, we find a $\xi = 2$ works better in matching with the benchmark Athena++ simulations. Hence, we use $\xi = \xi_{\rm \model} \equiv 2$ in \model runs,
\begin{alignat}{1}
    &v_{\rm turb, \texttt{grad},  cell, \rm \model} = V_{\rm cell}^{1/3}\sqrt{\frac{1}{\xi_{\rm \model}}\sum_{i, j}^3\left(\frac{\partial v_j}{\partial x_i}\right)^2}.
    \label{eq:final_mogli}
\end{alignat}
Later, in Sec.~\ref{sec:verify_kol} and \ref{sec:verify_local} we test our \model model with both Kolmogorov scaling-based ($\texttt{kol}$) and velocity gradient-based (\texttt{grad}) methods for local turbulent velocity estimation, respectively. See Appendix~\ref{app:turb_2d} for more general version Eq.~\eqref{eq:final_mogli}, which can also be applied to 2D geometries.
\begin{figure}
    \includegraphics[width=\columnwidth]{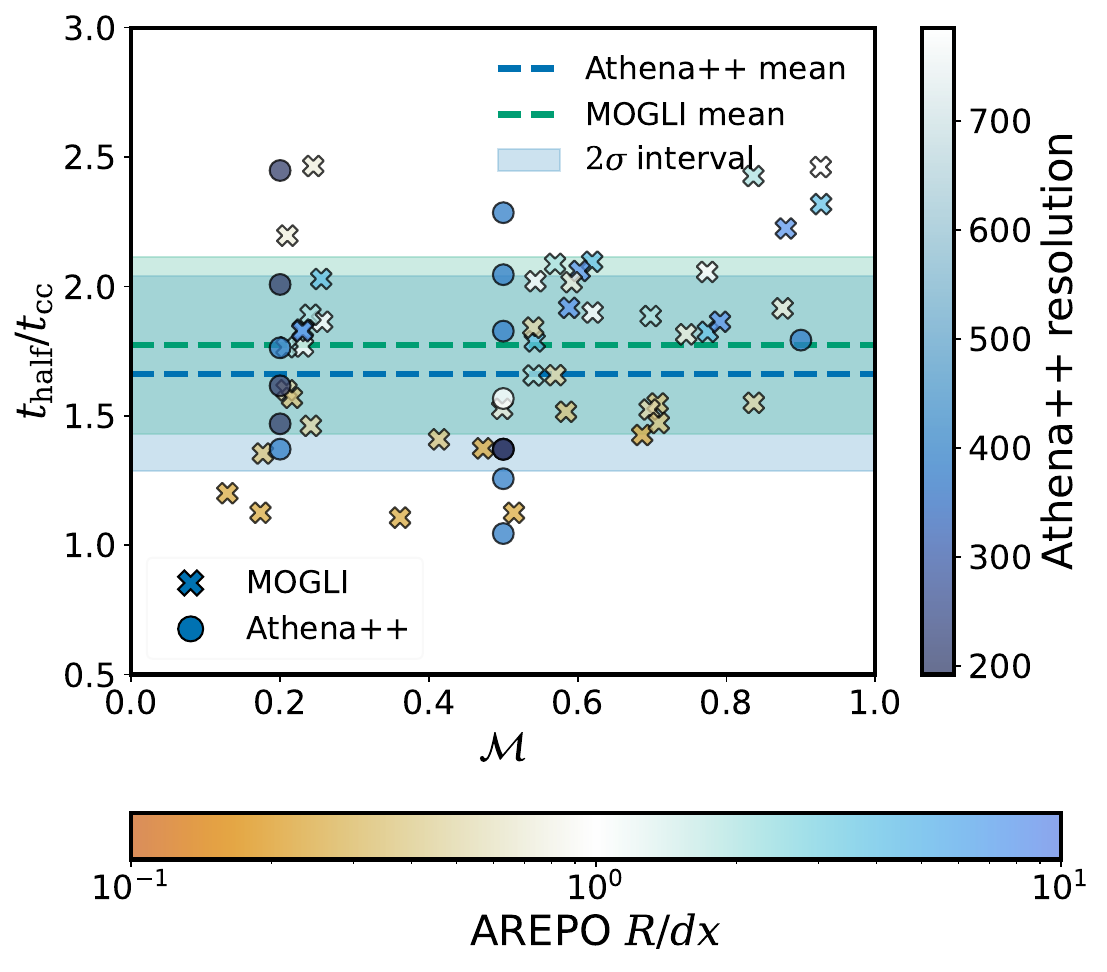}
    \caption{Scatter plot of the half mass time ($t_{\rm half}$) normalised to the initial cloud-crushing timescale ($t_{\rm cc} = \chi R_{\rm cloud}/v_{\rm turb}$), for different turbulent Mach numbers. \texttt{Athena++} simulations with different resolutions ($192^3$, $384^3$, and $768^3$, represented by the colour of the point) and turbulence random seeds to capture the inherent stochasticity of cold gas destruction in a turbulent medium.}
    \label{fig:half_mass_time}
\end{figure}

\begin{figure*}
    \includegraphics[width=\textwidth]{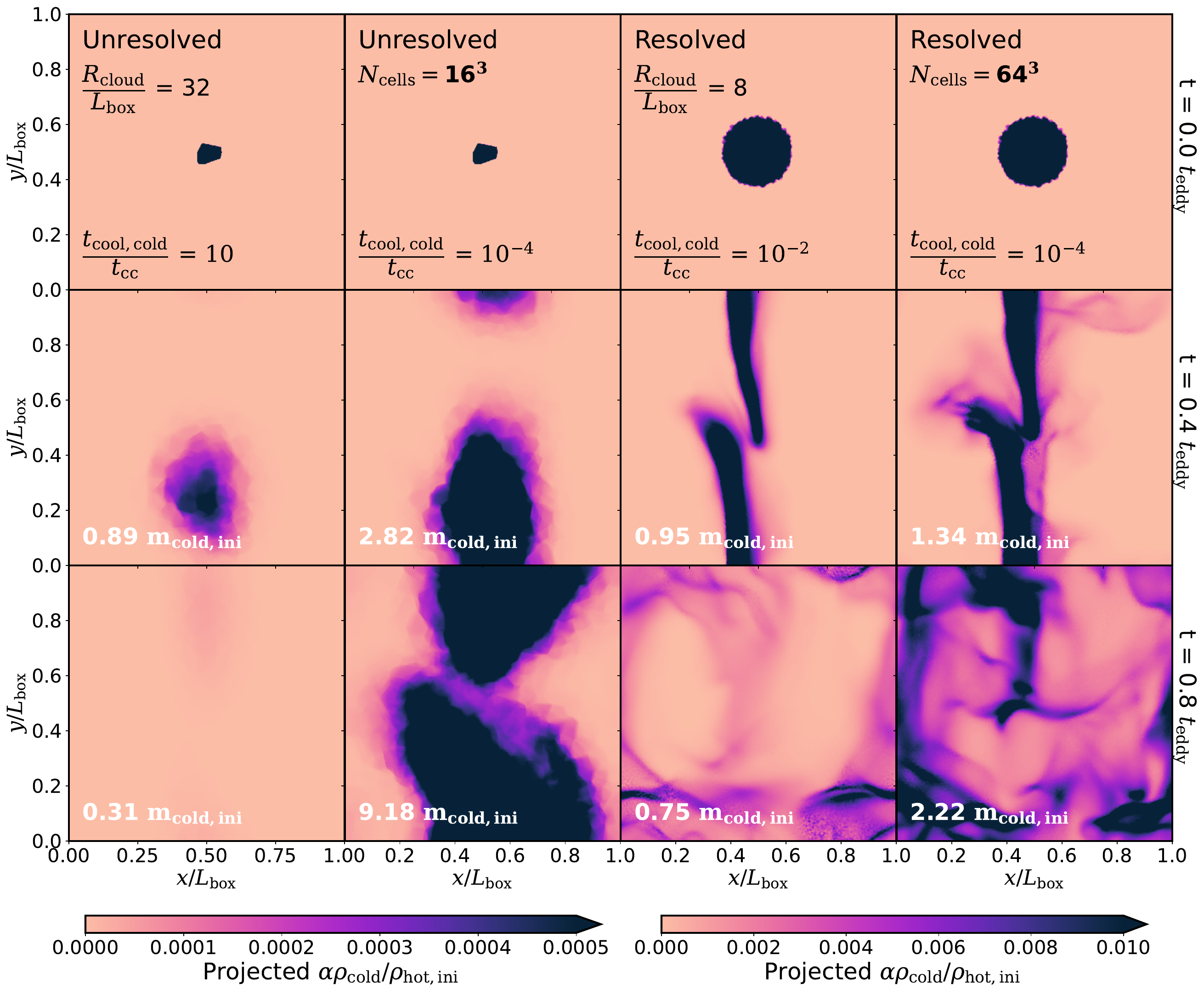}
    \caption{Projected $\alpha\rho_{\rm cold}/\rho_{\rm hot, ini}$, i.e. $\int_{\rm los} \alpha (\rho_{\rm cold}/\rho_{\rm hot, ini}) dz/L_{\rm box}$, plots at different times for \model runs with $\mathcal{M} = 0.5$ different $t_{\rm cool, cold}/t_{\rm cc}$ values. Two columns on the left show the evolution of an unresolved ($L_{\rm box}/R_{\rm cloud} = 32$) initial cloud and two columns on the right show the evolution of a resolved initial ($L_{\rm box}/R_{\rm cloud} = 32$) cold cloud for destruction and survival regimes. We find that the clouds with short cooling timescales, i.e. $t_{\rm cool, cold} = \{10, 10^{-2}\}t_{\rm cc}$ survive and grow, while clouds with long cooling timescales, i.e. $t_{\rm cool, cold} = 10^{-4}t_{\rm cc}$ end up losing cold gas and get destroyed, as expected from the results of previous studies \citep{GronkeTurb2022}.}
    \label{fig:overview_kol}
\end{figure*}

\subsection{Cold gas surface and cross-sectional area}
\label{subsec:area_factors}
Both the mass exchange from hot to cold and cold to hot medium naturally depend on the exposure of the two media, and thus crucially depend on the size of the interface area in any given cell. In order to account for non-spherical and overlapping cold gas structures, we use two area factors in the model introduced in Sec.~\ref{subsec:mixing}~\&~\ref{subsec:grow} (cf. Eq.~\eqref{eq:mdot_mix}~\&~\eqref{eq:mdot_grow}).

The first area factor, $2h(\alpha)$, corresponds to the dependence on the interface area between the two fluids. The second area factor, $A_{\rm cross}(\alpha)$, is for the cross-sectional area along the relative velocity of the fluid.

We take a Monte-Carlo approach to approximate these area factors. We generate a fixed grid box with a varying number of equally sized, spheres, allowing for overlapping spheres\footnote{We also tried different sphere size distributions but did not find any major impact on the results while significantly increasing the number of free parameters.}. We calculate the surface area ($A$) of the resulting ensemble of spheres using the \texttt{scikit-learn}'s marching cubes algorithm \citep{scikit-learn, MarchingCube}, and volume fraction ($\alpha$) as the ratio of the volume of encapsulated cells and grid volume ($V_{\rm box}$). We repeat this exercise with different radii of spheres ($R_{\rm sphere}$).

\begin{figure*}
    \includegraphics[width=\textwidth]{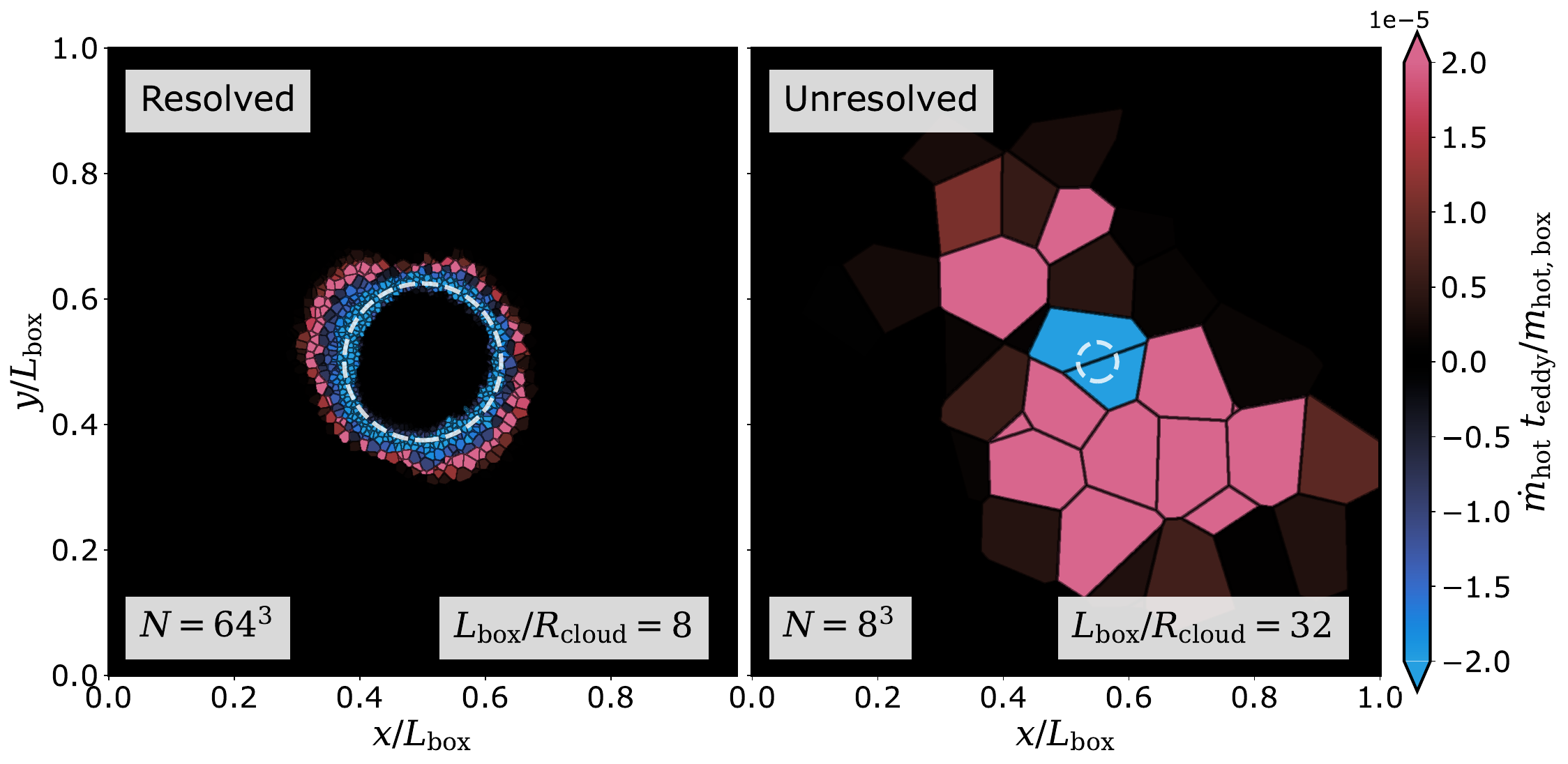}
    \caption{Early-time hot fluid mass flux ($\dot{m}_{\rm hot}$) slices, normalised with ratio of total hot fluid mass and eddy-turnover time ($m_{\rm hot, box}/t_{\rm eddy}$), for \model simulations with resolved and unresolved cold gas clouds, at $\mathcal{M} = 0.5$. The left panel shows an example of a resolved cold gas cloud with $64^3$ cells and $L_{\rm box}/R_{\rm cloud} = 8$, where the cloud is bigger than the grid cells and grid cells inside the volume of the cloud have an $\alpha = 1-\alpha_{\rm floor}$. On the other hand, the right panel shows the slice for \model simulation with an unresolved cold gas cloud, with $8^3$ cells and $L_{\rm box}/R_{\rm cloud} = 32$. This shows how the model is able to distinguish between the interior and exterior of the resolved cloud and the mass exchange only occurs at the interface around the cloud. The dashed circles show the corresponding initial cold gas cloud size in the simulations.}
    \label{fig:mass_flux_kol}
\end{figure*}

We find, as shown in Fig.~\ref{fig:ar_vs_volfrac}, that the quantity $AR_{\rm sphere}/(2V_{\rm box})$ seems to follow a general relation with the volume fraction.
\begin{alignat}{2}
    &\frac{AR_{\rm sphere}}{2V_{\rm box}} &&= \alpha h(\alpha) =
    \begin{cases}
        \alpha & \text{if } \alpha \in [0, 0.4) \\
        0.4 &  \text{if } \alpha \in [0.4, 0.8) \\
        2-2\alpha &  \text{if} \alpha \in [0.8, 1]
    \end{cases}
    \label{eq:interface_area}
\end{alignat}
where $h(\alpha)$ is a function of volume fraction, which we define for convenience in later derivations\footnote{We also tested similar fits to the points in Fig.~\ref{fig:ar_vs_volfrac}, including higher order ones, but it does not change the results significantly.}.

\begin{figure}
    \includegraphics[width=\columnwidth]{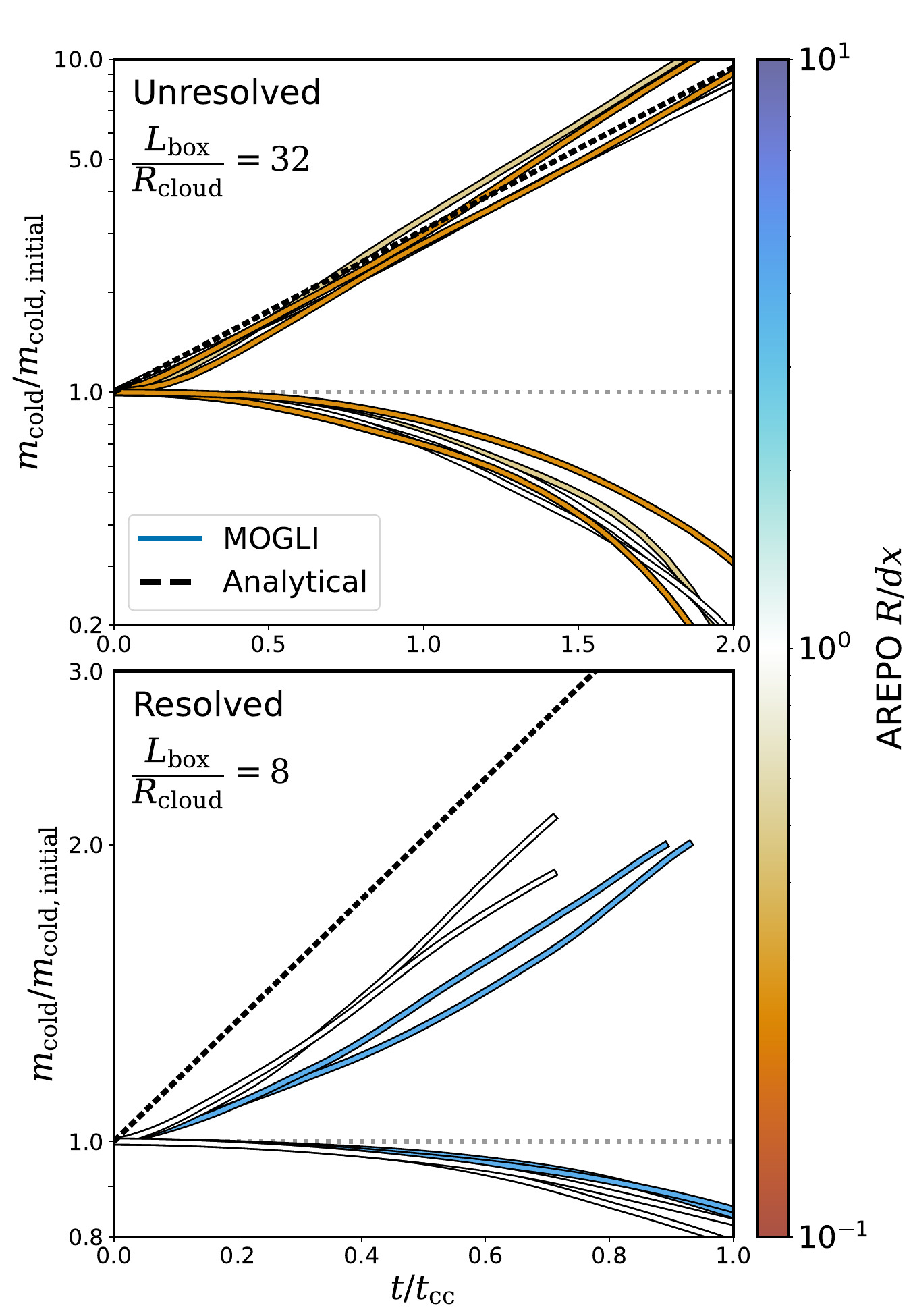}
    \caption{Cold gas evolution in MOGLI runs with time, normalised to the initial cloud-crushing time ($t_{\rm cc}$), with $\mathcal{M}_{\rm turb} = 0.5$. The two groups of curves correspond to $t_{\rm cool, cold}/t_{\rm cc} = \{10^{-4}, 10\}$. The solid lines show the cold gas evolution, as the total mass of the cold fluid, with the colour of the line denoting the initial $R_{\rm cloud}/dx$. \textit{Top panel} shows the evolution for simulations with unresolved initial cloud $L_{\rm box}/R_{\rm cloud} = 32$ and \textit{bottom panel} shows the same for resolved initial cloud $L_{\rm box}/R_{\rm cloud} = 8$. The black dashed line shows the expected exponential growth of the simulations which grow, with the growth time ($t_{\rm grow}$) calculated using Eq.~\eqref{eq:tgrow}. We find a good agreement between the analytically expected growth rates and \model runs.}
    \label{fig:coldmass_kol}
\end{figure}

Assuming that the spheres are composed of cold gas, with an overdensity of $\chi$ and density of $\chi\rho_{\rm hot}$, we obtain a relation for the interface area as a function of the volume fraction,
\begin{alignat}{3}
    & A &&= \frac{2h(\alpha)\alpha V_{\rm box}}{R_{\rm sphere}} && = \frac{2h(\alpha)m_{\rm cold}}{\chi\rho_{\rm hot} R_{\rm sphere}}.
    \label{eq:area_alpha}
\end{alignat}
Substituting the area ($A$) from Eq.~\eqref{eq:area_alpha} in the expression for mass exchange  (e.g., \citealp{Tan2021RadiativeCombustion}) we obtain
\begin{alignat}{1}
    &\dot{m}_{\rm cold} = A\rho_{\rm hot} v_{\rm flux} =  \frac{ m_{\rm cold}}{\chi R_{\rm sphere}/ v_{\rm flux}} 2h(\alpha) = 2h(\alpha)  \frac{m_{\rm cold}}{t_{\rm flux}}
    \label{eq:area_factor_h}
\end{alignat}
which introduces the cold gas surface area factor of $2h(\alpha)$. Here, $t_{\rm flux}\sim\chi R_{\rm sphere}/v_{\rm flux}$ is the naive mass doubling/halving time of a spherical object. Note that the factor of $2$ along $h(\alpha)$ can explain the fudge factor $0.5$ used in $t_{\rm grow}$ in \citep{GronkeTurb2022}. At low cold gas volume fractions, Eq.~\eqref{eq:area_factor_h} reduces to the empirical exponential growth rate expression observed in previous studies like \citep{GronkeTurb2022}.

\begin{figure}
    \includegraphics[width=\columnwidth]{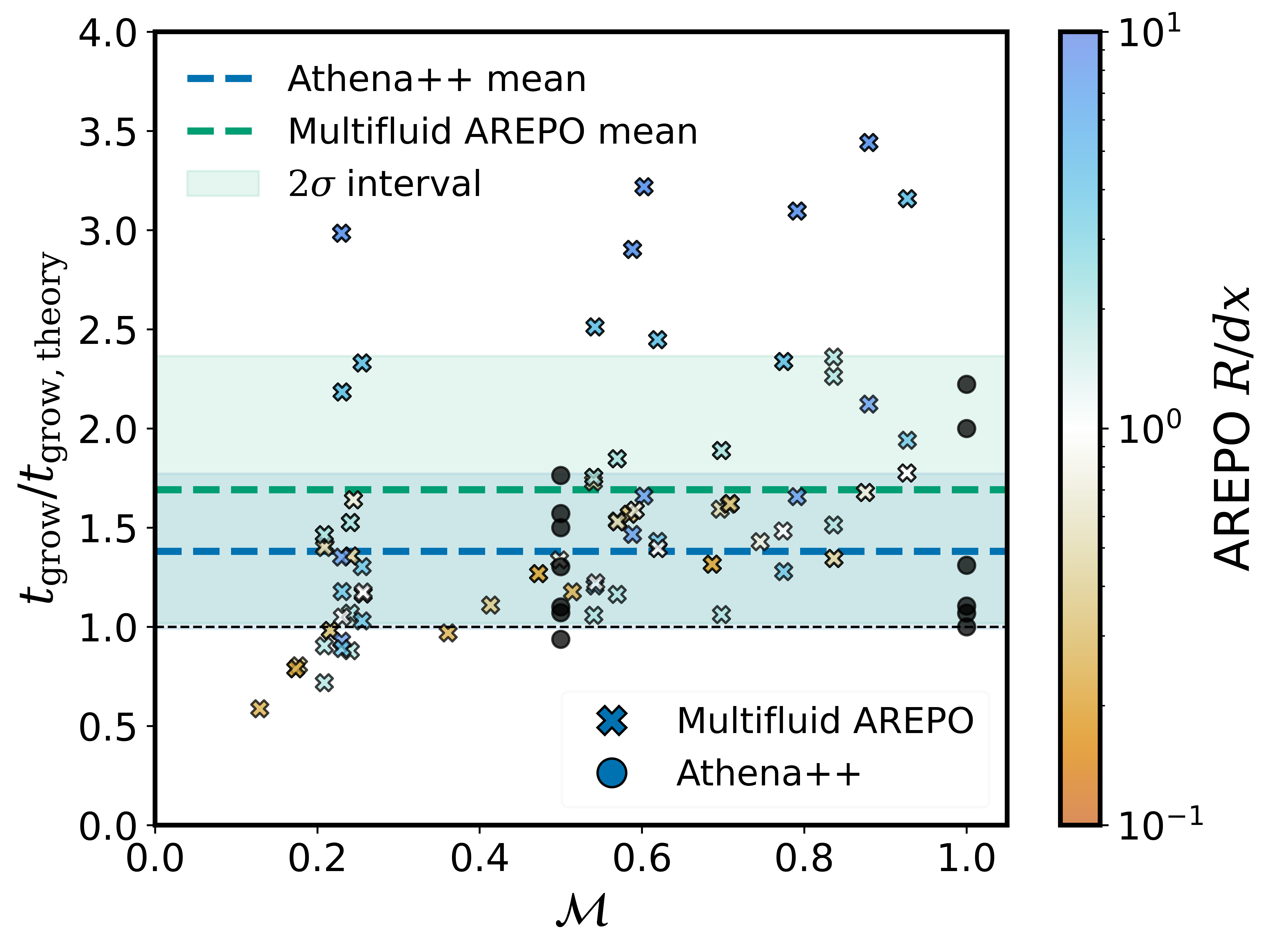}
    \caption{Scatter plot of the ratio of the $t_{\rm grow}$ from the simulations and the analytical $t_{\rm grow, theory}$ (Eq.~\eqref{eq:tgrow}), across different turbulent Mach number ($\mathcal{M}$). Crosses show the values from the \model runs, with the colours denoting their "Resolvedness" ($R/dx$), while the black circles show the values from benchmark \texttt{Athena++}. The set of points for benchmark \texttt{Athena++} include values calculated from simulations from \citet{GronkeTurb2022, Das2024}. We also show the means as dashed lines and $2\sigma$ intervals as shaded regions of the benchmark \texttt{Athena++} and \model runs. The comparison shows only a marginal difference between the benchmark \texttt{Athena++} and \model runs in the means with significant overlap between the scatter.}
    \label{fig:double_mass_time}
\end{figure}

We take a similar approach to approximate the dependence of cross-sectional area in the direction of relative velocity, on the volume fraction. Assuming an isotropic distribution of the spheres, the cross-sectional area should be independent of the direction of the relative velocity. We calculate the cross-sectional area ($A_{\rm cross}$) and volume fraction ($\alpha$) of a varying number of randomly distributed spheres along the three axes of the grid and repeat it for different radii of the spheres ($R_{\rm sphere}$). Fig.~\ref{fig:cross_section_area} shows the dependence of $A_{\rm cross}$ normalised to the grid's cross-sectional area, i.e. $A_{\rm box}=L_{\rm box}^2$, on the volume fraction of the spheres. There is a scatter, where larger individual spheres lead to a lower cross-sectional area for the same volume fraction. However, the trend converges with smaller individual clouds, and we use a sigmoid function to approximate this relation.
\begin{alignat}{2}
    &A_{\rm cross}(\alpha) &&= V_{\rm cell}^{2/3}\left(\frac{e^{10 \alpha} - 1}{e^{10 \alpha} + 1} \right)\label{eq:A_cross}
\end{alignat}
With this, we have the complete and fully-defined \model model and can move to verifying this model in the next section.

\section{Verification: With Kolmogorov turbulence estimation}
\label{sec:verify_kol}

As the turbulent velocity estimation is a fundamental part of the model, we verify the two estimates separately and present the results in two sections. In this section, we test \model with the Kolmogorov scaling-based turbulent velocity estimate. We use high-resolution \texttt{Athena++} turbulent box simulations with resolved initial cold gas cloud as the benchmark to compare the multifluid simulations with the \model subgrid model. For a model to be declared operational, we require the multifluid simulation to agree with the benchmark \texttt{Athena++} simulations, even if the
\begin{itemize}
    \item initial cold gas clump is resolved or unresolved, i.e., $R_{\rm cloud}/dx>1$ or $<1$, respectively.\footnote{Note that throughout the text, we use $R_{\rm cloud}$ refers to the size of the initial cold gas cloud. $l_{\rm cold}$ in the previous section refers to the cold gas length scale inside a particular grid cell.},
    \item resolution of the simulation is varied ($dx$), regardless of the ``resolvedness'' of the cloud,
    \item turbulent velocity is changed ($\mathcal{M}$),
    \item random driving of the turbulence is different.
\end{itemize}

In the next sections, we test the different parts of the model across this parameter space.
First, we only add the contributions from drag and turbulent mixing of cold gas and verify this reduced version of \model with the benchmark \texttt{Athena++} simulations in Sec.~\ref{subsec:ver_adiab_mix}. In Sec.~\ref{subsec:ver_rad_mix}, we verify the full \model model again with small-scale high-resolution \texttt{Athena++} simulations and results.

\subsection{Non-radiative Mixing}
\label{subsec:ver_adiab_mix}

We start with testing a reduced version of the \model model. This is to independently verify the different parts of the model. For the first set of tests, we only include the contributions from $\Qdrag$ and $\Qmix$, i.e. $\Qdot_{\rm non-rad} = \Qdrag + \Qmix$. This setup is analogous to the turbulent mixing of cold and hot gas in a non-radiative box, i.e. without radiative cooling.

We run the \texttt{Athena++} turbulent box simulations and introduce a cold cloud after driving the turbulence, without radiative cooling. In the absence of cooling of the mixed gas, inevitably, the cold cloud loses mass at an approximately exponential rate. As shown in Fig.~\ref{fig:cold_gas_adiab}, we find the same behaviour in the \model simulations with the subgrid model using $\Qdot_{\rm non-rad}$. We vary the $L_{\rm box}/R_{\rm cloud}$ in the range $[8, 64]$, resolution per direction in $[8, 64]$, turbulent Mach number, $\mathcal{M}$ in $[0.2, 0.75]$ and random seed for turbulence driving.

In \texttt{Athena++}, we define cold gas mass as the total mass of gas cells with temperature below $8\times 10^4$K and in \model, the cold gas mass is the total mass of the cold fluid in the box. We calculate this for the snapshots separately and find the time taken for the total cold gas mass to reach half its initial value, i.e. $t_{\rm half}$. Fig.~\ref{fig:half_mass_time} shows a scatter plot of the half mass time ($t_{\rm half}$) normalised to the initial cloud-crushing timescale ($t_{\rm cc} = \chi^{1/2} R_{\rm cloud}/v_{\rm turb}$)\footnote{Throughout the text, $t_{\rm destroy}$ refers to the cold gas destruction timescale within the cell (cf. Eq.~\ref{eq:t_destroy}), while $t_{\rm cc}$ refers to the cloud destruction timescale of the initial cold gas cloud.}, for different turbulent Mach numbers. As a benchmark, we use resolved \texttt{Athena++} simulations with different resolutions ($192^3$, $384^3$ and $768^3$, represented by the colour of the point) and turbulence random seeds to capture the inherent stochasticity of cold gas destruction in a turbulent medium. Due to this inherent stochasticity, we show the mean and $2\sigma$ interval around the mean for the benchmark \texttt{Athena++} and corresponding \model simulations. For the \model runs, the colour of the point in Fig.~\ref{fig:half_mass_time} shows how resolved or unresolved the initial cloud is via the ratio of the cloud radius and grid cell size ($R/dx$).

We find that the mean and scatter of the cold gas destruction from the subgrid model with the Kolmogorov turbulent velocity estimation agree well with the benchmark \texttt{Athena++} and both cluster with a factor of $\sim 2$ near the analytical value of $\sim t_{\rm cc}$. This verifies that the source function contribution for mixing, $\Qmix$ is working as expected and leads to a physically consistent behaviour.

\subsection{Radiative mixing}
\label{subsec:ver_rad_mix}

Next, we include the remaining source function contribution for the growth of cold gas via mixing, $\Qgrow$. This gives us the full subgrid model for radiative mixing with $\Qdot = \Qdrag + \Qmix + \Qgrow$. With the full subgrid model, the \model simulations are analogous to resolved turbulent box simulations with radiative cooling, similar to simulations in \citet{Das2024, GronkeTurb2022}.

We run the \model simulations with different turbulent Mach numbers ($\mathcal{M}$) in [0.2, 0.75], multiple resolutions per direction in [8, 64], two different random seeds for turbulent driving and different values of $t_{\rm cool, cold}/t_{\rm cc}$. We calculate the value of $t_{\rm cool, cold}$ from the $t_{\rm cc} = \chi^{1/2}R_{\rm cloud}/v_{\rm turb}$ and required value of their ratio. We use our \texttt{Athena++} runs with different resolutions, turbulent Mach numbers random turbulence seeds, and cold cloud sizes ($R_{\rm cloud}$) as the benchmark for comparison. This includes our simulations and the results from \citet{GronkeTurb2022, Das2024}.

\subsubsection{Morphology}
\label{subsubsec:ver_rad_mix_morph}

Fig.~\ref{fig:overview_kol} shows the 2D maps of projected $\alpha \rho_{\rm cold}$ ($\int_{\rm los} \alpha \rho_{\rm cold} dz/L_{\rm box}$) plots at different times, for \model runs with $\mathcal{M} = 0.5$ different $t_{\rm cool, cold}/t_{\rm cc}$ values. Two columns on the left show the evolution of an unresolved ($L_{\rm box}/R_{\rm cloud} = 32$) initial cloud and two columns on the right show the evolution of a resolved initial ($L_{\rm box}/R_{\rm cloud} = 32$) cold cloud for destruction and survival regimes. We find that the clouds for runs with short $t_{\rm cool, cold}$ show growth of cold gas, while the clouds in runs with longer $t_{\rm cool, cold}$ get destroyed, as expected from the results of previous studies \citep{GronkeTurb2022}. \\

Fig.~\ref{fig:mass_flux_kol} shows the early-time hot fluid mass flux ($\dot{m}_{\rm hot}$) slices, normalised with ratio of total hot fluid mass and eddy-turnover time ($m_{\rm hot, box}/t_{\rm eddy}$), for radiative \model simulations with resolved and unresolved cold gas clouds, at $\mathcal{M} = 0.5$. The left panel shows a case of a resolved cold gas cloud with $64^3$ cells and $L_{\rm box}/R_{\rm cloud} = 8$, where the cloud is bigger than the grid cells and grid cells inside the volume of the cloud have an $\alpha = 1-\alpha_{\rm floor}$.
It is clear that cells located within the cloud lose hot mass and gain cold mass ($\dot m_{\rm hot}<0$), while cold gas is being mixed into cells located just outside the cloud which thus gain hot mass and lose cold mass ($\dot m_{\rm hot} > 0$).

On the other hand, the right panel of Fig.~\ref{fig:mass_flux_kol} shows the slice from a \model simulation with an unresolved cold gas cloud, $8^3$ cells and $L_{\rm box}/R_{\rm cloud} = 32$. This shows how the model can distinguish between the interior and exterior of the resolved cloud and ensures that the mass exchange only occurs at the interface around the cloud. The dashed circles show the corresponding initial cold gas cloud size in the simulations.

\subsubsection{Cold gas growth and survival}
\label{subsubsec:ver_rad_mix_grow}

We use the same definition of cold gas as in Section~\ref{subsec:ver_adiab_mix}, for both benchmark \texttt{Athena++} runs and \model runs to calculate the cold gas mass in different snapshots and obtain the evolution of cold gas mass with time. Fig.~\ref{fig:coldmass_kol} shows the temporal evolution of the cold gas, normalised to the initial cold gas mass, in \model runs with $\mathcal{M} = 0.5$. We include simulations with two different cold cloud sizes ($L_{\rm box}/R_{\rm cloud}$), two values for $t_{\rm cool, cold}/t_{\rm cc} = \{10^{-4},\,10\}$, two values for turbulence random seed and multiple resolutions. We only show the evolution till $m_{\rm cold, box} < 0.9 m_{\rm total, box}$, as beyond that point the growth of the cold gas stagnates due to the deficiency of hot gas in the box and the theoretical predictions from previous studies do not apply anymore. The colour of the lines shows how "Resolved" ($R/dx>1$) or "Unresolved"($R/dx<1$) the initial cold gas cloud is. We find that the \model runs, regardless of the varied degrees of resolutions, also show growth of cold gas when $t_{\rm cool, cold} \ll t_{\rm cc}$ and destruction when $t_{\rm cool, cold} \gg t_{\rm cc}$, in agreement with the previous results using resolved single-fluid simulations \citep{GronkeTurb2022, Das2024}.

In Fig.~\ref{fig:coldmass_kol}, we also show the corresponding analytical, cold gas growth curve as black dashed lines, and the \model runs agree well with the predicted exponential growth curve. The analytical growth timescale, $t_{\rm grow, theory}$, is given by \citep{GronkeTurb2022}
\begin{alignat}{2}
    &\dfrac{t_{\rm grow, theory}}{t_{\rm cc}} &&= 0.5~\chi \left( \dfrac{t_{\rm cool, cold}}{t_{\rm cc}} \right)^{1/2} \left( \dfrac{L_{\rm box}}{R_{\rm cloud}} \right)^{1/6} .
    \label{eq:tgrow}
\end{alignat}
Note that while this equation is of the same form as the implemented Eq.~\eqref{eq:tgrow_implemented}, on a per-cell basis, it is important to point out that whether our simulations recover the correct global growth rate is far from obvious.

Next, we quantitatively compare the cold gas growth rates between the benchmark \texttt{Athena++} and the \model runs. We calculate the cold gas mass doubling time and use it to calculate the $t_{\rm grow}$. Fig.~\ref{fig:double_mass_time} shows the ratio of the $t_{\rm grow}$ from the simulations and the analytical $t_{\rm grow, theory}$ from Eq.~\eqref{eq:tgrow}, across different turbulent Mach number ($\mathcal{M}$). Crosses denote the values from the \model runs, with the colours denoting their "Resolvedness" ($R/dx$), while the black circles show the values from benchmark \texttt{Athena++}. The set of points for benchmark \texttt{Athena++} include values of data from \citet{GronkeTurb2022, Das2024}. We also show the mean values as dashed lines and $2\sigma$ intervals as shaded regions, in blue for the benchmark \texttt{Athena++} and in green for \model runs. The comparison shows only a marginal difference in the means with significant overlap between the scatter of values from benchmark \texttt{Athena++} and \model runs. This verifies that the \model subgrid model accurately captures the growth of cold gas to a reasonable extent.

\begin{figure}
    \includegraphics[width=\columnwidth]{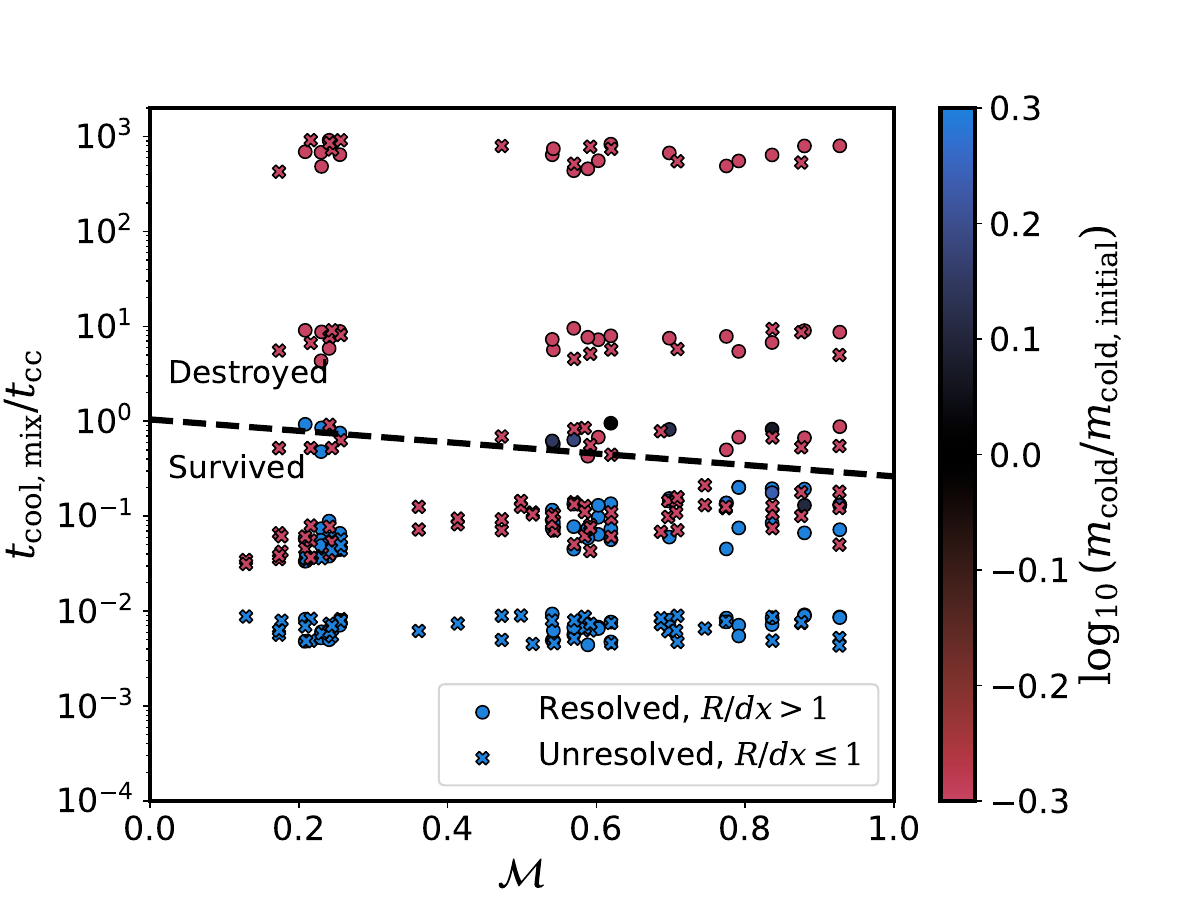}
    \caption{Scatter plot of survival or destruction of cold gas in the \model runs, in a parameter space of $t_{\rm cool, mix}/t_{\rm cc}$ and turbulent Mach number, $\mathcal{M}$, where $t_{\rm cool, mix}$ (c.f. Eq.~\eqref{eq:tcoolmix}). The circles show the points from resolved ($R/dx>1$) \model simulations, while crosses denote the unresolved ($R/dx<1$) simulations. The colour of the points denotes the ratio of total final cold fluid mass, averaged over the last 10 snapshots, normalised to the initial value. The black dashed line shows the survival criterion from \citet{GronkeTurb2022}, and we find that \model can reproduce this survival criterion as an emergent behaviour. Note that the points are randomly shifted vertically by a factor of $1.5$ for clarity.}
    \label{fig:survival_kol}
\end{figure}
Apart from growth rates, a working subgrid model should consistently agree with the survival criterion of cold gas clouds obtained in previous studies using resolved single-fluid simulations \citep{GronkeTurb2022}. In Fig.~\ref{fig:survival_kol}, we reproduce the survival plot from \citet{GronkeTurb2022} with \model runs. It shows the survival or destruction of cold gas in the \model runs, in a parameter space of $t_{\rm cool, mix}/t_{\rm cc}$ and turbulent Mach number ($\mathcal{M}$), where $t_{\rm cool, mix}$ refers to the cooling time of the intermediate mixed gas,
\begin{alignat}{2}
 & t_{\rm cool, mix} &&= t_{\rm cool}(T_{\rm mix}, \rho_{\rm mix})
 \label{eq:tcoolmix}
\end{alignat}
where the $T_{\rm mix}=\sqrt{T_{\rm cold}T_{\rm hot}}$ and $\rho_{\rm mix} = \sqrt{\rho_{\rm cold}\rho_{\rm hot}}$ are the temperature and density of the intermediate mixed gas \citep{Begelman1990}.

\begin{figure}
    \includegraphics[width=\columnwidth]{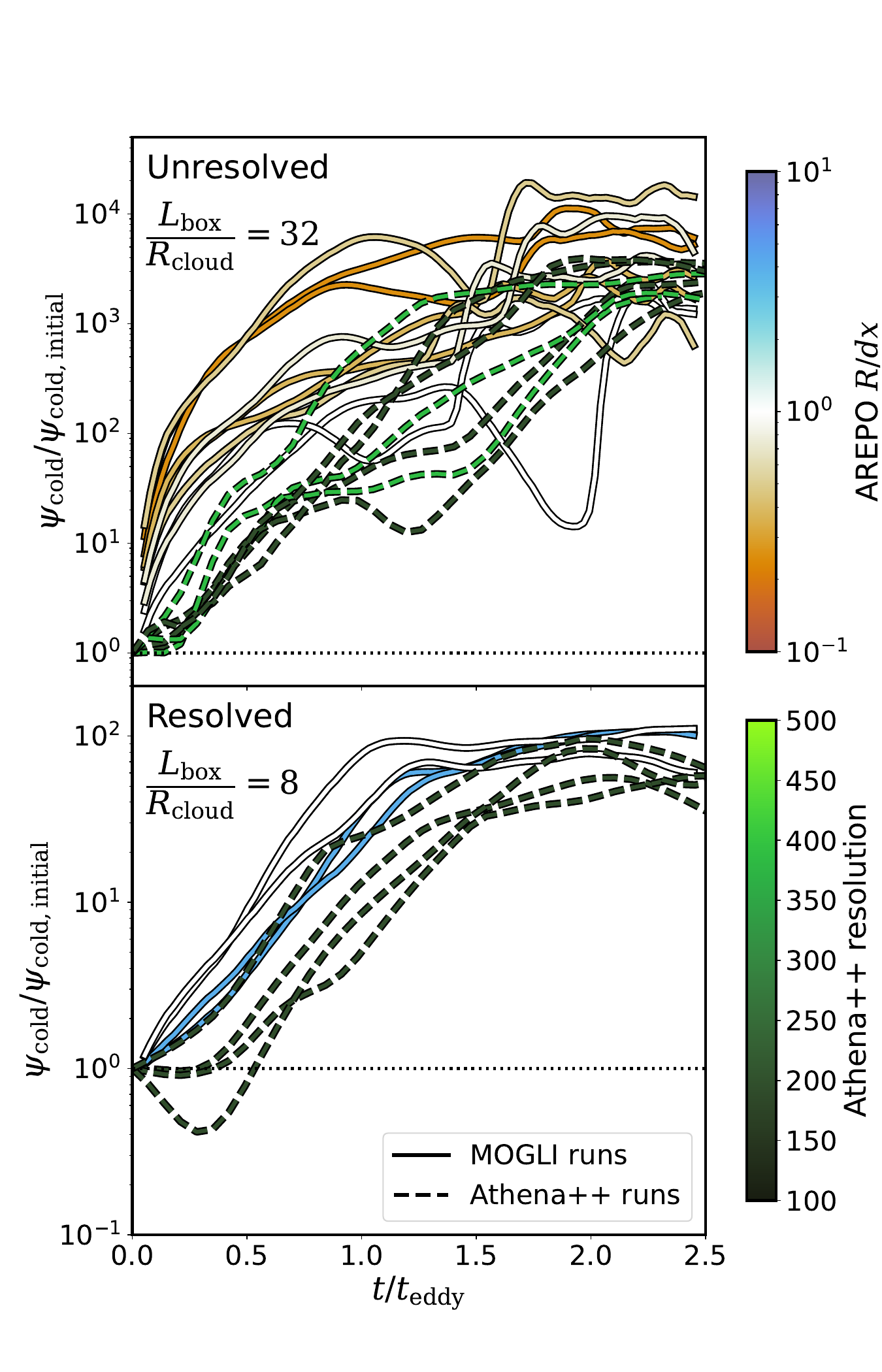}
    \caption{Evolution of the cold gas dispersion, normalised to its initial value, in the benchmark \texttt{Athena++}, as dashed lines, and \model runs, as solid lines, with time normalised with turbulent eddy turnover time. The colour of dashed lines shows the resolution of the \texttt{Athena++} simulations, while the colour of the solid lines shows the ``resolvedness'' of the initial cold cloud in the \model runs, i.e. $R/dx$. \textit{Left panel} shows the evolution of \model runs with resolved initial clouds at $L_{\rm box}/R_{\rm cloud}=8$, and the corresponding \texttt{Athena++} runs. \textit{Right panel} shows the same but for \model runs with unresolved initial cloud at $L_{\rm box}/R_{\rm cloud}=32$, and the \texttt{Athena++} runs.}
    \label{fig:disp_kol}
\end{figure}
As the \model runs only use the cold gas cooling time ($t_{\rm cool, cold}$), unlike our \texttt{Athena++} runs which use a full cooling function, we assume the CIE cooling curve from \citet{Wiersma2009} (as used in the benchmark \texttt{Athena++} runs) to evaluate $t_{\rm cool, mix}=\chi t_{\rm cool, cold}\Lambda(T_{\rm mix}) / \Lambda(T_{\rm cold})$. For our choice of $T_{\rm cold}$, $\chi$ this results in $t_{\rm cool,mix}\approx 64 t_{\rm cool,cold}$

The circles in Fig.\ref{fig:survival_kol} denote runs with ``resolved'' ($R/dx>1$) initial cold clouds and crosses represent the ``unresolved'' ($R/dx\leq1$) initial cold clouds and the colour of the points show the final cold gas mass, averaged over the last 10 snapshots and normalised with the initial cold gas mass. We show the survival criterion found in \citet{GronkeTurb2022} and confirmed in \citet{Das2024} as a dashed black line along with annotation for the survival and destruction regimes. As there are a large number of points with very similar parameters, we randomly displace the points vertically by a factor of $1.5$, for clarity. We find that the \model runs agree well with the survival criteria. We also see some stochasticity in simulations that lie close to the criterion. This is an expected behaviour seen in resolved single-fluid simulations due to the inherent randomness of turbulence. This verifies the ability of the subgrid model to accurately reproduce the survival and destruction of cold gas as an emergent behaviour.

\begin{figure}
    \includegraphics[width=\columnwidth]{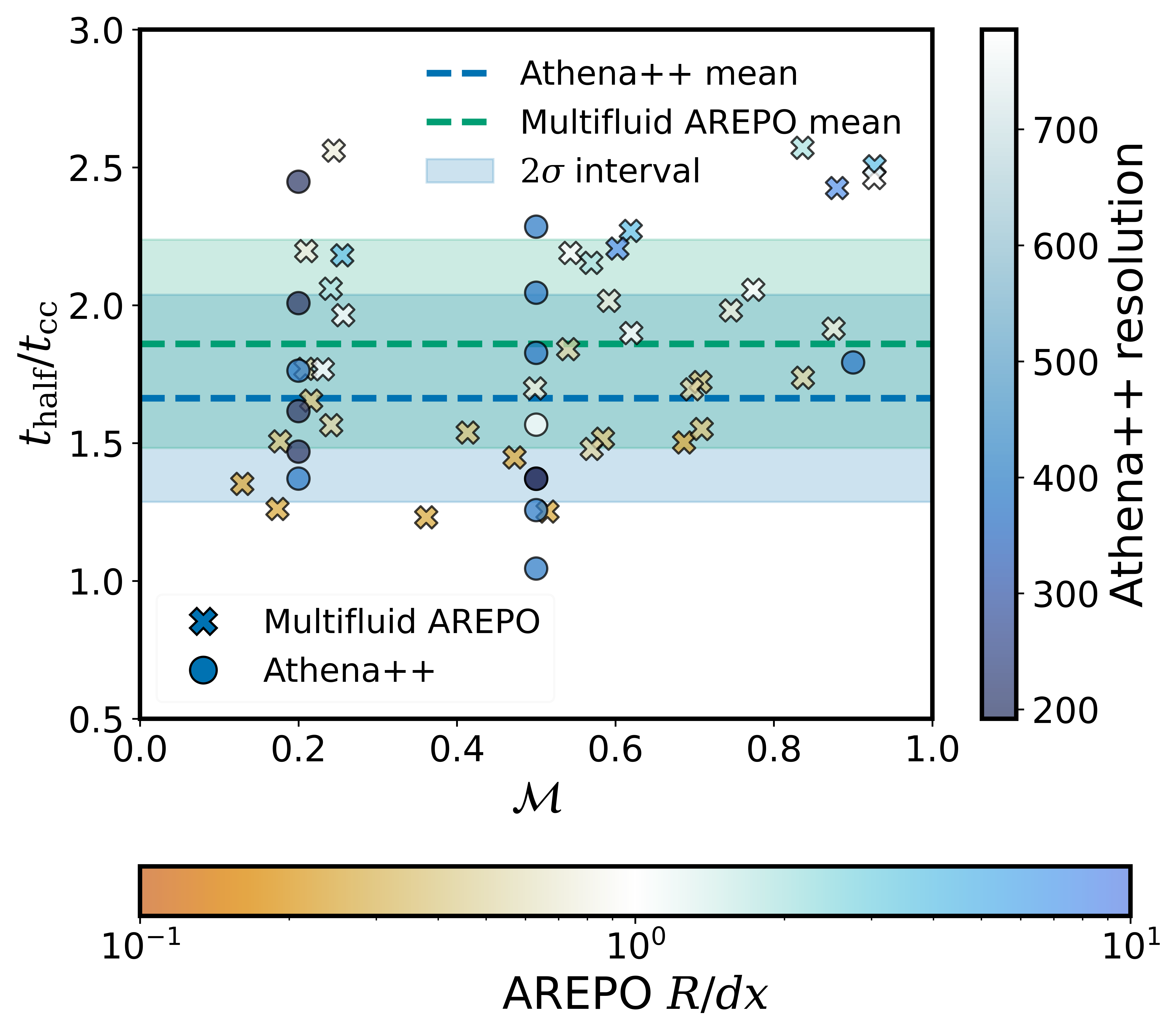}
    \caption{Same as Fig.~\ref{fig:half_mass_time} but with the gradient-based local turbulence estimation. Scatter plot of the half mass time ($t_{\rm half}$) normalised to the initial cloud-crushing timescale ($t_{\rm cc} = \chi R_{\rm cloud}/v_{\rm turb}$), for different turbulent Mach numbers. \texttt{Athena++} simulations with different resolutions ($192^3$, $384^3$, and $768^3$, represented by the colour of the point) and turbulence random seeds to capture the inherent stochasticity of cold gas destruction in a turbulent medium.}
    \label{fig:half_mass_time_local}
\end{figure}
\begin{figure}
    \includegraphics[width=\columnwidth]{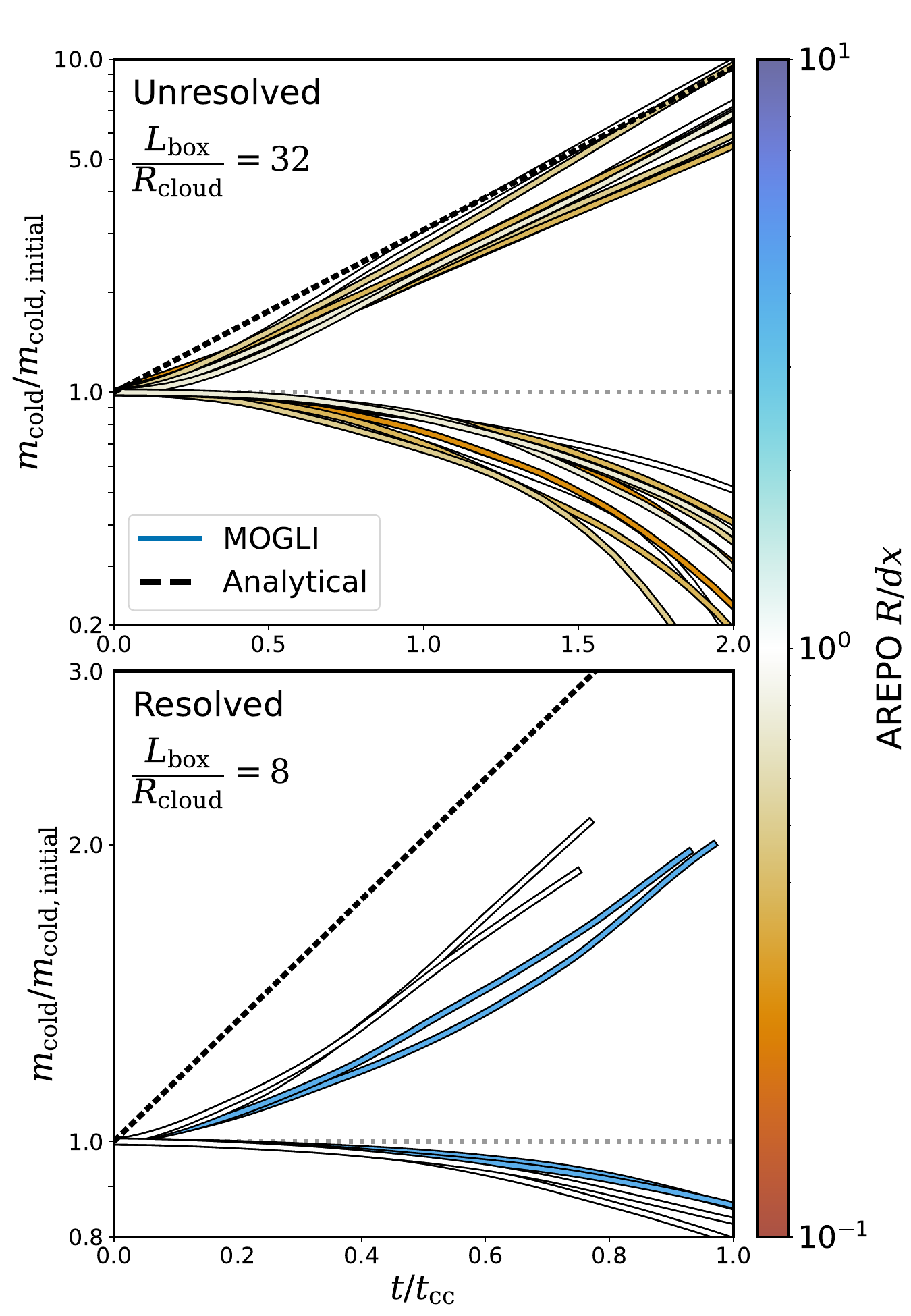}
    \caption{Same as Fig.~\ref{fig:coldmass_kol} but with the gradient-based local turbulence estimation. Cold gas evolution in MOGLI runs with time, normalised to the initial cloud-crushing time ($t_{\rm cc}$), with $\mathcal{M}_{\rm turb} = 0.5$. The two groups of curves correspond to two different values of $t_{\rm cool, cold}/t_{\rm cc}= \{10^{-4}, 10\}$. The solid lines show the cold gas evolution, as the total mass of the cold fluid, with the colour of the line denoting the initial $R_{\rm cloud}/dx$. \textit{Top panel} shows the evolution for simulations with unresolved initial cloud $L_{\rm box}/R_{\rm cloud} = 32$ and \textit{bottom panel} shows the same for resolved initial cloud $L_{\rm box}/R_{\rm cloud} = 8$. The black dashed line shows the expected exponential growth of the simulations which grow, with the growth time ($t_{\rm grow}$) calculated using Eq.~\eqref{eq:tgrow}. We find a good agreement between the analytically expected growth rates and \model runs.}
    \label{fig:coldmass_local}
\end{figure}
\begin{figure}
    \includegraphics[width=\columnwidth]{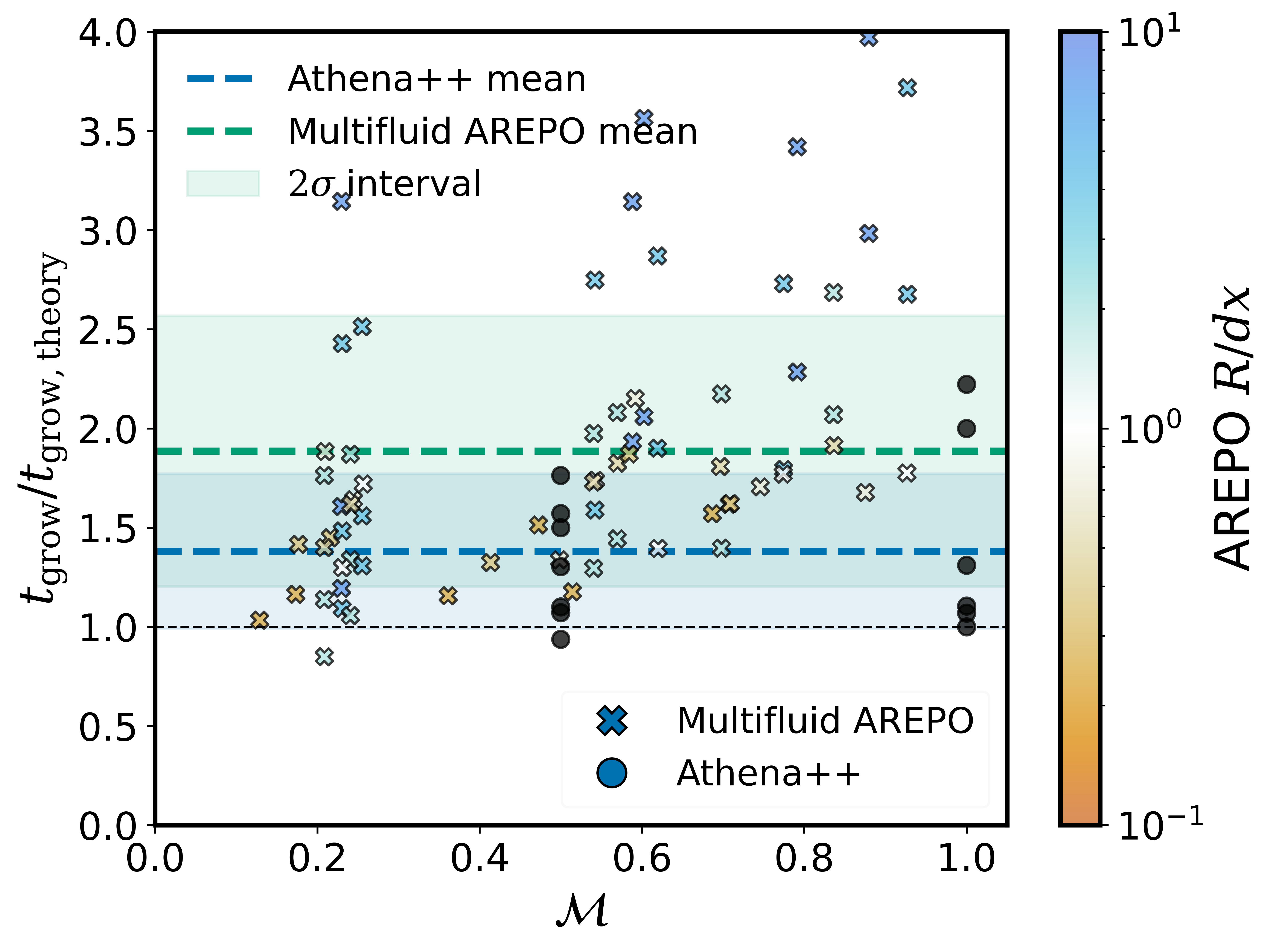}
    \caption{Same as Fig.~\ref{fig:double_mass_time} but with the gradient-based turbulence estimation. Scatter plot of the ratio of the $t_{\rm grow}$ from the simulations and the analytical $t_{\rm grow, theory}$ (Eq.~\ref{eq:tgrow}), across different turbulent Mach number ($\mathcal{M}$). Crosses show the values from the \model runs, with the colours denoting their "Resolvedness" ($R/dx$), while the black circles show the values from benchmark \texttt{Athena++}. The set of points for benchmark \texttt{Athena++} include values calculated from simulations from \citet{GronkeTurb2022, Das2024}. We also show the means as dashed lines and $2\sigma$ intervals as shaded regions of the benchmark \texttt{Athena++} and \model runs. The comparison shows only a marginal difference between the benchmark \texttt{Athena++} and \model runs in the means with significant overlap between the scatter.}
    \label{fig:double_mass_time_local}
\end{figure}

\subsubsection{Cold gas dispersion}
\label{subsubsec:ver_rad_mix_disp}
As the simulation evolves, the cold gas is expected to get progressively more dispersed with time and can be an important mechanism for the transport of cold gas in large-scale simulations. We test for differences in the dispersion of cold gas in the benchmark \texttt{Athena++} and \model simulations. We define a quantity, $\psi$, as a proxy for the dispersion of cold gas,
\begin{alignat}{2}
    &\psi_{\rm cold} &&= \prod_{i=[1,2,3]} \left[\dfrac{\sum_{\rm cold} m_{\rm cold}~\delta x_{i, \rm com}}{\sum_{\rm cold} m_{\rm cold}} \right]
    \label{eq:disp}
\end{alignat}
where, $\delta x_{i, \rm com} = |x_{i, \rm cell} - x_{i, \rm com, cold}|$ is the distance, along $i^{\rm th}$ axis, between the cold gas/fluid centre-of-mass and cell centre, in a periodic box. We run benchmark \texttt{Athena++} and the corresponding \model runs with initial $R_{\rm cloud}/l_{\rm shatter} = 310$, varying resolution and different turbulent random seed. We repeat this calculation on the snapshots from \texttt{Athena++} and \model runs to obtain the temporal evolution of the cold gas dispersion ($\psi_{\rm cold}$). Fig.~\ref{fig:disp_kol} shows the temporal evolution of the $\psi_{\rm cold}$, normalised to its initial value, in the benchmark \texttt{Athena++} as dashed lines, and \model runs as solid lines. The colour of dashed lines shows the resolution of the \texttt{Athena++} simulations, while the colour of the solid lines shows the ``resolvedness'' of the initial cold cloud in the \model runs, i.e. initial $R/dx$. The top panel shows the evolution of \model runs with unresolved initial clouds at $L_{\rm box}/R_{\rm cloud}=32$, and the corresponding \texttt{Athena++} runs. The bottom panel shows the same but for \model runs with unresolved initial cloud at $L_{\rm box}/R_{\rm cloud}=8$, and the \texttt{Athena++} runs.

In general, in both ``resolved'' and ``unresolved'' cloud cases, while the cold gas dispersion is higher in \model runs, the growth of dispersion follows qualitatively similar evolution with similar timescales. The higher dispersion is likely due to the poorer resolution for the highly unresolved, i.e. low $R/dx$, which leads to higher numerical diffusion in volume fraction. This can be improved in future with a higher-order solver in the multifluid code, but the differences fall within an acceptable range for the current study.

With the above tests, we verify that the \model model, with the Kolmogorov turbulence estimation, can capture all the different facets of cold gas behaviour, i.e. the survival, destruction rates, growth rates and cold gas dispersion in a turbulent medium, in an emergent way. In the next section, we repeat the verification tests from this section for \model model with our new gradient-based local turbulence estimation method (\texttt{grad}).

\section{Verification: With Velocity gradient-based turbulence estimation}
\label{sec:verify_local}

As the Kolmogorov estimation method (\texttt{kol}) assumes isotropic and steady-state turbulence over the simulation domain, it cannot be applied in setups with evolving or strongly varying turbulence. For more general scenarios with temporal and/or spatial variation in turbulence, we need a method to estimate the local turbulent velocity from the local fluid properties. We implement the \texttt{grad} method for local turbulence estimation based on velocity gradients to get around this limitation of the \texttt{kol} method. As described in Sec.~\ref{subsubsec:grad_vturb}, we use the local velocity gradient calculated using neighbouring cells in \texttt{AREPO} \citep{Pakmor2016} to estimate the velocity dispersion in the neighbourhood. Once we know the turbulent velocity in this neighbourhood, we scale it to the cold gas scales by assuming a fully developed Kolmogorov turbulence below those grid scales.
\begin{figure}
    \includegraphics[width=\columnwidth]{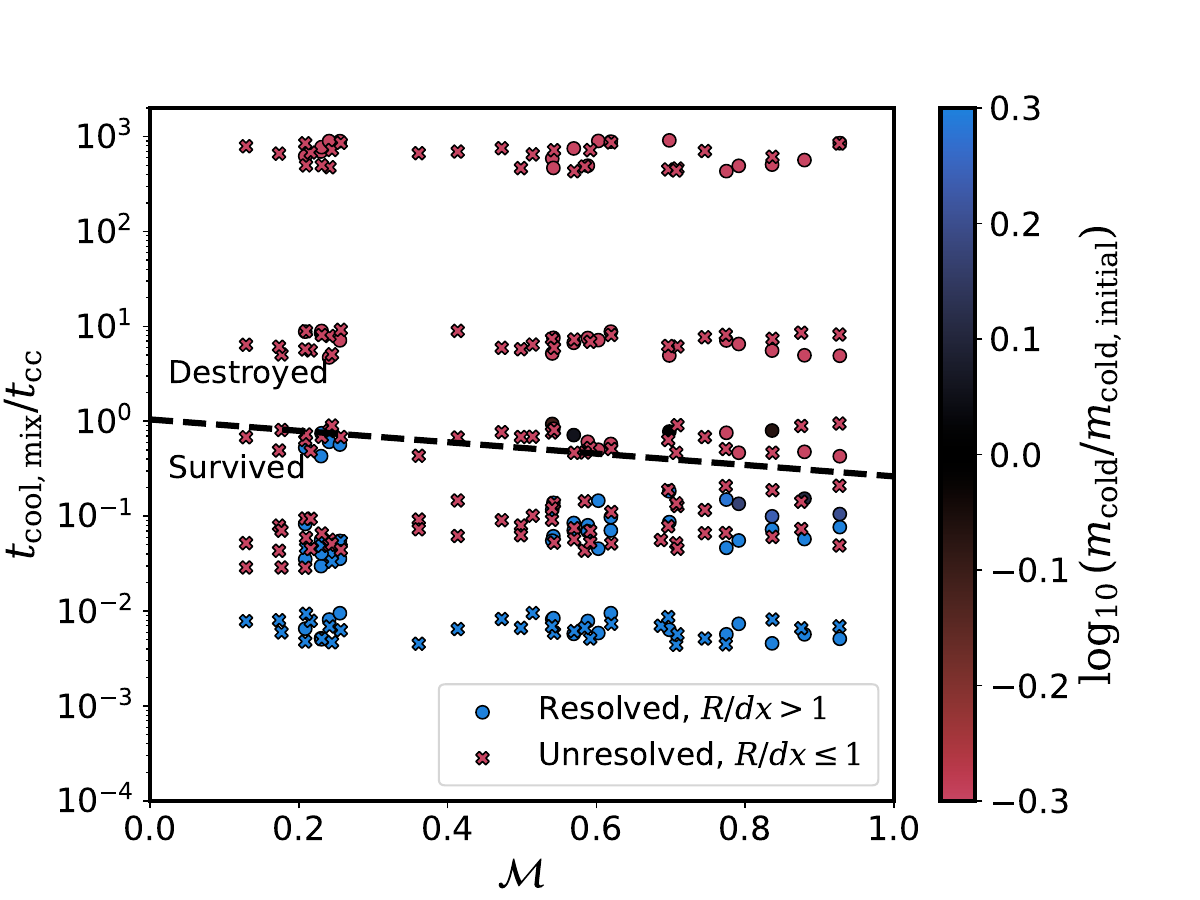}
    \caption{Same as Fig.~\ref{fig:survival_kol} but with the gradient-based turbulence estimation. Scatter plot of survival or destruction of cold gas in the \model runs, in a parameter space of $t_{\rm cool, mix}/t_{\rm cc}$ and turbulent Mach number, $\mathcal{M}$, where $t_{\rm cool, mix}$ (c.f. Eq.~\eqref{eq:tcoolmix}). The circles show the points from resolved ($R/dx>1$) \model simulations, while crosses denote the unresolved ($R/dx<1$) simulations. The colour of the points denotes the ratio of total final cold fluid mass, averaged over the last 10 snapshots, normalised to the initial value. The black dashed line shows the survival criterion from \citet{GronkeTurb2022}, and we find that \model can reproduce this survival criterion as an emergent behaviour. Note that the points are randomly shifted vertically by a factor of $1.5$ for clarity.}
    \label{fig:survival_grad}
\end{figure}

We repeat our verification tests from Section~\ref{sec:verify_kol} to confirm the accuracy of \model with the gradient-based turbulent velocity estimation. We start with checking the non-radiative turbulent mixing rates. Fig.~\ref{fig:half_mass_time_local}, analogous to Fig.~\ref{fig:half_mass_time}, shows the cold gas half time for non-radiative \model runs ($\Qdot = \Qdrag +\Qmix$) with the \texttt{grad} method. For the \texttt{grad} method, we find a slightly higher separation between the mean of the benchmark \texttt{Athena++} and \model runs, and a wider scatter in both cases when compared to the \texttt{kol} method of estimation. The differences are minimal and within acceptable limits. So, we can conclude that the \texttt{grad} method works well for non-radiative mixing.

For the full model, i.e., including radiative mixing, first, we check the evolution of cold gas mass evolution in Fig.~\ref{fig:coldmass_local}, analogous to Fig.~\ref{fig:coldmass_kol}. It shows the same behaviour where, regardless of the resolution, the clouds with faster cooling time, i.e. $t_{\rm cool, cold}/t_{\rm cc} \ll 1$, grow at expected rates while the clouds with slow cooling, i.e. $t_{\rm cool, cold}/t_{\rm cc} \gg 1$, get destroyed. Subsequently, we repeat the quantitative verification of the growth rates in Fig.~\ref{fig:double_mass_time_local}, similar to the comparison made in Fig.~\ref{fig:double_mass_time}. We find marginally higher differences in the mean value of $t_{\rm grow}/t_{\rm grow, theory}$ (cf. Sec.~\ref{subsubsec:ver_rad_mix_grow}) of the benchmark \texttt{Athena++} runs and \model runs, in comparison to the differences with \texttt{kol} method, while the standard deviation is similar.

Next, we verify the emergent survival criterion from \model with the \texttt{grad} estimation. In Fig.~\ref{fig:survival_grad}, we show the survival and destruction of the cold cloud, analogous to Fig.~\ref{fig:survival_kol}. We find that the \model runs still agree well with the survival criterion from \citet{GronkeTurb2022}, albeit slightly worse than the \texttt{kol} method.

\begin{figure}
    \includegraphics[width=\columnwidth]{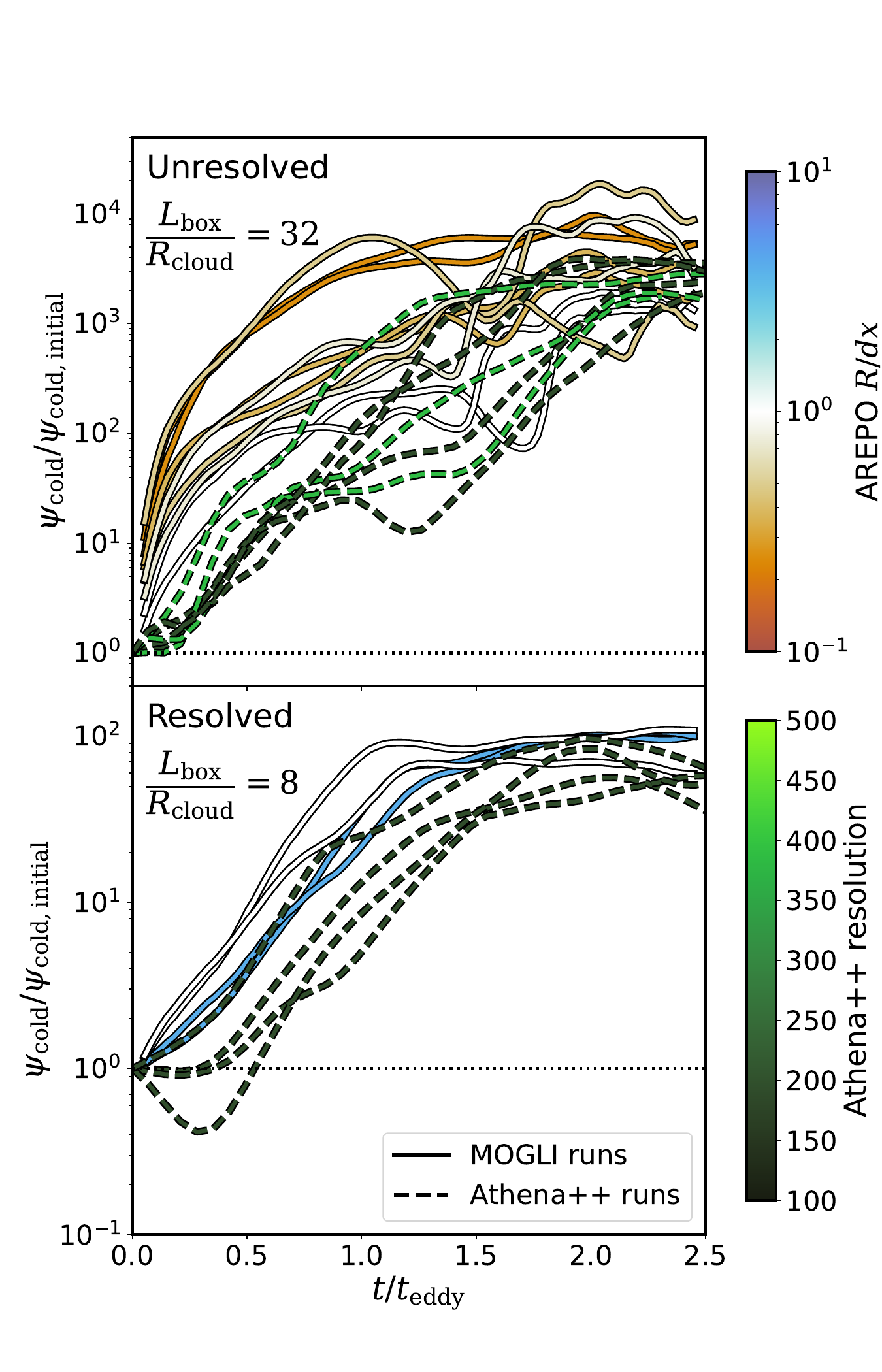}
    \caption{Same as Fig.~\ref{fig:disp_kol} but with gradient-based turbulence estimation. Evolution of the cold gas dispersion, normalised to its initial value, in the benchmark \texttt{Athena++}, as dashed lines, and \model runs, as solid lines, with time normalised with turbulent eddy turnover time. The colour of dashed lines shows the resolution of the \texttt{Athena++} simulations, while the colour of the solid lines shows the ``resolvedness'' of the initial cold cloud in the \model runs, i.e. $R/dx$. \textit{Top panel} shows the same but for \model runs with unresolved initial cloud at $L_{\rm box}/R_{\rm cloud}=32$, and the \texttt{Athena++} runs. \textit{Bottom panel} shows the evolution of \model runs with resolved initial clouds at $L_{\rm box}/R_{\rm cloud}=8$, and the corresponding Athena++ runs. }
    \label{fig:disp_sgst}
\end{figure}

Finally, we check for any differences in the dispersion of cold fluid using the local turbulent estimator. Fig.~\ref{fig:disp_sgst} shows the evolution of cold gas dispersion $\psi_{\rm cold}$ (cf. Eq.~\ref{eq:disp} in Sec.~\ref{subsubsec:ver_rad_mix_disp}), analogous to Fig.~\ref{fig:disp_kol} for \texttt{kol} method. We find that the cold gas dispersion follows an almost identical evolution with \texttt{grad} method, in Fig.~\ref{fig:disp_sgst} as with \texttt{grad} in Fig.~\ref{fig:disp_kol}.

With the series of tests above, we can conclude that \model model with the velocity gradient-based local turbulence estimation (\texttt{grad}) method also accurately captures the subgrid behaviour of cold gas with a good agreement with the benchmark \texttt{Athena++} and analogous \model runs with the \texttt{kol} simulations.
Since the average turbulence estimates are consistent and the agreement between \model and the resolved \texttt{Athena++} runs was demonstrated in the previous section, this outcome is unsurprising. However, this assumption does not hold for simulations focused on local measures (e.g., local dispersion) or where gas flows vary significantly, making a `global' concept of turbulence inapplicable. In summary, we find the \texttt{grad} method to be more suitable for most astrophysical applications and thus adopt it as the default in \model. Nonetheless, the initial comparison with the simpler \texttt{kol} method provides a valuable opportunity to isolate and understand the underlying physical mechanisms.

\section{Discussion}
\label{sec:discussion}
\subsection{Model showcase}
\label{sec:showcase}

\begin{figure*}
    \includegraphics[width=\textwidth]{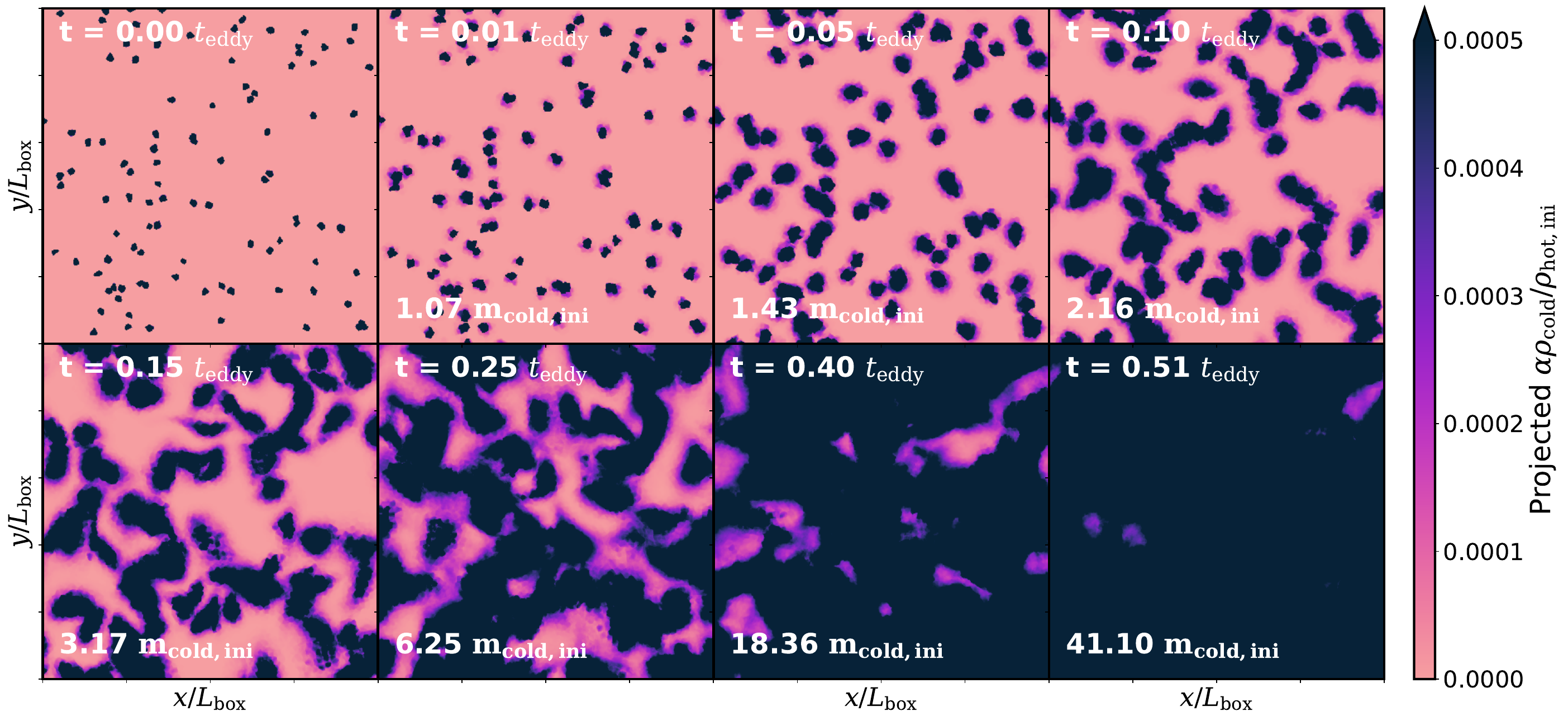}
    \caption{Evolution of projected $\alpha\rho_{\rm cold}/\rho_{\rm hot, ini}$ for a turbulent box with \model, \texttt{grad} method, $64^3$ cells, 100 unresolved clouds with a radius $L_{\rm box}/256$, where $L_{\rm box}$ is the box size, and $t_{\rm cool}/t_{\rm cc} = 5\times10^{-4}$. The unresolved clouds grow and subsequently fill the box due to its finite size. We will need a box with $\sim 3000^3$ cells to run an analogous simulation in a single-fluid code without a subgrid model.}
    \label{fig:showcase}
\end{figure*}

After the testing and verification of the \model model in Sec.~\ref{sec:verify_kol}~\&~\ref{sec:verify_local}, we can use it to simulate a setup to show the strengths of such a model. We simulate a turbulent box with $64^3$ cells and $\mathcal{M} = 0.5$, and introduce 100 unresolved clouds with a radius of $L_{\rm box}/256$, where $L_{\rm box}$ is the box size. To run an analogous setup in a single-fluid simulation code without a subgrid model, we need to resolve the individual clouds with around 10 cells across their radius, hence requiring a single-fluid simulation with $\sim 3000^3$ cells. Such single-fluid simulations require a considerable computational cost, while our analogous \model multifluid simulation can capture the relevant cold gas evolution with just a $64^3$ cells simulation with negligible computational cost. Fig.~\ref{fig:showcase} shows the evolution projected $\alpha \rho_{\rm cold}/\rho_{\rm hot, ini}$ for the showcase simulation using our \model model with the \texttt{grad} local turbulence estimation method.

This shows the potentially massive savings in computational times with the \model model, over a brute force method of resolving the cold gas structures in a single-fluid code without a subgrid model allowing for configurations infeasible with traditional single-fluid simulations. We plan to study several such scenarios in future work.

\subsection{The need for a multiphase subgrid model}
\label{sec:discu_need}
With the ISM, CGM, and ICM, multiphase media are ubiquitous in and around galaxies and, thus, crucial to our understanding of galaxy formation and growth.
Not only is the new fuel for star-formation channelled through the multiphase galactic halos \citep[see reviews by][]{Donahue2022,Faucher2023} but gas is also expelled from galaxies through multiphase galactic winds \citep[][]{Veilleux2020, Heckman2017arXiv}. Even the gas within the galaxy is highly multiphase with a much wider range of temperatures, from $\sim 10^2$K gas in molecular gas to $>10^7$K gas in supernova ejecta. Thus, correctly modelling multiphase gas dynamics is the foundation of accurately modelling gas transport, its conversion into stars and their subsequent impact on the galaxy when they end in supernovae and lead to feedback. Hence, understanding multiphase gas is a cornerstone of understanding the wider baryon cycle.

While great strides are being made to model the interstellar medium including its multiphase nature correctly in dedicated, small domain simulations \citep{Walch2015SILCC, Kim2017TIGRESS}, this is much more problematic in a larger scale, cosmological simulations. Nevertheless, modern large-scale cosmological simulations using `zoom in' and adaptive techniques, based on targeting a mass resolution, manage to reach a parsec-scale resolution in certain dense regions within the ISM while at the same time capturing the effects of the larger environment.

However, the same techniques cannot be applied to the halo of galaxies spanning $\sim 100\,$kpc in radius (for a Milky Way-sized galaxy at $z\sim 0$; \citealp{Tumlinson2017TheMedium}). In these large volumes, recent observations constrain the upper limit to the size of this cold gas to be $\lesssim 10\,$pc \citep{Lan2017ApJ...850..156L,Crighton2015Metal-enrichedGalaxy,Schaye2007MNRAS.379.1169S,Rauch1999ApJ...515..500R}.
This implies that simulation of a single halo would require at least $(100,000)^3$ resolution elements -- a task infeasible even for the next generation of supercomputers. For example, modern cosmological simulations require tens of millions of CPU hours to simulate $\sim 10^9$ particles/cells (e.g., \texttt{Thesan}, \citealp{Kannan2022}). Extrapolating this figure to $(100,000)^3$ cells, we would need $\sim10^{7}$ million core hours. Thus, even an optimally parallelised run on one million cores (approximately the number of CPU cores on the exascale supercomputer El Capitan) such a run would take $10^7$ hours, i.e. $\sim 1000$ years. Of course, this is a simplified approximation; however, it still illustrates the point that even next-generation exascale machines will not be able to resolve the CGM structure observed to date.

The inability of modern cosmological simulations to achieve the resolution needed to model the (observed) cold gas structures found in the halos directly comes from the fact they are non-converged in even the most basic cold gas properties such as the amount of $\sim 10^4\,$K gas found \citep[see extensive discussion of this in the literature, e.g.,][]{Faucher2023,vandeVoort2021,Hummels2024ApJ...972..148H, Ramesh2024MNRAS.528.3320R, Peeples2019ApJ...873..129P}. This implies that when comparing such simulations to CGM observables, it is impossible to know whether a (mis)match is due to the physics implemented in this simulation (e.g., the feedback mechanisms, and star formation models) or if this is a transient agreement that would vanish with a higher resolution.

In addition, while the simulation projects mentioned above can significantly increase the number of resolution elements in the CGM, it is important to note that all these projects start with cosmological zoom simulations of individual halos in which memory usage is not as much of an issue. However, to study cold gas microphysics across the galaxy population, this technique would have to be applied to a large number of halos, posing significant additional requirements on the available memory. Since modern uniform mass resolution galaxy formation simulations are limited not solely by compute resources but also by memory, such an undertaking would only allow very limited increases in the number of resolution elements per halo, with no prospects to resolve the required spatial scales any time soon.

This limitation due to a highly multiscale problem is not new to astrophysics. For about two decades now, we have known that supernovae feedback -- taking place on initially minuscule scales -- is crucial to shaping galaxy morphology \citep[e.g.][]{NaabOstriker}. The way these indispensable `feedback' processes are modelled in astrophysical simulations is through a `subgrid model', that is, by including their effects on larger scales through (empirical) source and sink terms.

Our model \model is exactly such a subgrid model but for a two-phase multifluid medium. While it is not the first model of this kind (cf. \S~\ref{sec:previous_work} for a comparison to previous work), it includes \textit{(i)} a solid Eularian-Eularian numerical framework common in engineering \citep{Weinberger2023} implemented in the popular code \texttt{AREPO} \citep{Rainer2020AREPOPublic, Springel2010MNRASArepo}, and \textit{(ii)} coupling terms between the phases based on and verified with small-scale simulations and combustion inspired theory \citep{Tan2021RadiativeCombustion,GronkeTurb2022}.

This solid foundation implies that the number of free parameters in our model is at a bare minimum, i.e. it includes only two free parameters. Only in Eq.~\eqref{eq:mdot_mix}, the $\dot{m}_{\rm cold \rightarrow hot}$ includes a fudge parameter for the $\alpha_{\rm mass}$ threshold, and in Eq.~\eqref{eq:final_mogli}, $v_{\rm turb, \texttt{grad}, \model}$ includes $\xi_{\rm \model}$ (which is within a factor of $\sim 2$ of the theoretically expected value).
Such a low number of free parameters is important, in general, for a subgrid model to preserve the predictive power of the simulations -- but especially for a multiphase subgrid model such as \model, as all cells are affected by it. Thus, allowing for more degrees of freedom by, for example, altering directly the amount of cold gas found in the halo would diminish the predictive power rapidly.

Naturally more work is required (cf. \S~\ref{sec:future}) but this work is a strive in the right direction to ultimately model the multiphase gas in and around galaxies in a converged manner.

\subsection{Previous work}
\label{sec:previous_work}
As multiphase media are common in terrestrial applications, numerical tools to model them are extensively developed and widely used in various engineering disciplines, such as chemical, mechanical, and civil engineering. For example, in chemical engineering, they are critical for simulating processes like distillation, fluidized bed reactors, and mixing in multiphase reactors. In mechanical engineering, they enable the design and optimization of cooling systems involving liquid-vapor interactions, such as in heat exchangers and condensers. In civil engineering, they play a key role in understanding and managing subsurface flows, including groundwater contamination, oil recovery, and sediment transport in rivers \citep{Prosperetti_Tryggvason_2007}.

In astrophysics, however, dedicated methods to model multiphase flows are less used -- with some notable exceptions. \citet{Semelin2002}, for instance, already included a `warm' ($>10^4\,$K) and a `cold', pressureless phase as two separate fluids in their SPH scheme (alongside stars and dark matter) with the colder one being modelled via a `sticky particle scheme' \citep{Levinson1981ApJ...245..465L}, thus, allowing the particles to inelastically collide with each other. They include the possibility for `warm' particles to cool down and turn into `cold' ones, and vice-versa to evaporate `cold' particles due to supernovae feedback.
Using this scheme, \citet{Semelin2002} successfully perform a simulation of an isolated disk galaxy and study its evolution. While their source/sink terms are relatively large-scale and simplified, their approach represents a pioneering step toward incorporating multiphase processes in galaxy modelling (also see \citealp{Berczik2003} for a similar implementation or \citealp{Harfst2004} for an extension to three phases).

A similar approach is followed by \citet{Scannapieco2006}, who instead of using two different particle types in their SPH scheme, `decouple' particles with low entropy (essentially corresponding to a cold phase), thus, allowing neighbouring particles with different thermal properties. This implementation is meant to address the `overcooling problem' where the energy of a supernova ejected in the dense medium is instantly lost \citep[e.g.][]{Kim2015ApJ...802...99K, Smith2018MNRAS.478..302S}.
While the multiphase implementation of \citet{Scannapieco2006} is not specifically targeting galactic winds, their supernovae feedback implementation results in a separation of nearby particles into `hot' and `cold' ones, and, thus leads to an efficient wind launching. Their algorithm can thus be seen to be more in line with the more recent developments of SN ejected winds where particles are also decoupled \citep{SpringelHernquist2003MNRAS.339..289S, Okamoto2005MNRAS.363.1299O, Oppenheimer2006MNRAS.373.1265O}.

More recently, the multiphase nature of galactic winds has led to the development of several Eulerian-Lagrangian subgrid models where the cold phase is represented by particles within a hot gas cell \citep{Huang2020,Smith2024}. In these models, physically motivated source and sink terms between the phases akin to the ones used in this study have been implemented. However, as the particles representing the cold phase have zero extents, reproducing the dispersion of the cold medium (cf. \S~\ref{subsubsec:ver_rad_mix_disp}) would require particle splitting/merging which is currently not implemented in these models. 

Our work can be seen as a continuation of these efforts. While the previous astrophysical work mentioned above has focused on a colder phase represented via a particle, we have created a two-phase (a $\sim 10^4\,$K and a `hot' $\gtrsim 10^6\,$K phase) Eulerian-Eulerian subgrid model using the multifluid implementation of \citet{Weinberger2023} and coupling terms inspired by combustion theory which was validated previously using a range of small-scale numerical simulations \citep{Ji2019SimulationsLayers,Tan2021RadiativeCombustion,GronkeTurb2022,Tan2023Infalling}.
This implementation has the advantage that it can especially capture multiphase, turbulent media -- which are ubiquitously found in astrophysics, such as in the CGM, ICM and galactic winds -- as well as the cold gas disperses naturally.

In this regard, our approach is similar to the two-fluid model implemented in \texttt{ENZO} by \citet{2024MNRAS.535.1672B} during the work on this model. Note, however, that our cold fluid is represented as a compressible second fluid while their cold phase is assumed to be pressureless. While both should yield the same result in the case of vanishing volume filling fractions and temperature of the cold phase, a pressureless fluid can only represent unresolved cold clouds. The transition to marginally resolved or even fully resolved clouds where the pressure of the cold fluid displaces the hot fluid, i.e. the main focus of this work, cannot be modelled in the pressureless approach.
This also results in differences in source terms: their coupling terms between the phases are inspired by the classical `cloud crushing' problem, i.e., the interaction of a laminar hot wind and a cold cloud \citep[e.g.,][]{Klein1994} with the individual cold gas cloudlets radii being $r_{\mathrm{cl}}\approx \mathrm{min}(c_{\mathrm{s}}t_{\mathrm{cool}})$ inspired by the `shattering' scenario \citep{McCourt2018AGas}. In contrast, the presented work allows full flexibility in the respective volume filling fractions and allows, specifically, also the case where the typical length and timescales of unresolved and resolved processes are comparable. Thus, we needed to generalise the source terms to account for the finite availability of hot gas, of `shielding' of inner layers of resolved cold clouds and geometric considerations of the change in surface area when a substantial fraction of a cell is filled with cold gas and generally cannot assume an instantaneous equilibration of the cold gas into a universal cloud radius or cloud mass function.

\subsection{Limitations and future directions}
\label{sec:limitations} \label{sec:future}

Although our model \model takes a step in the direction of capturing the rich physics of small-scale cold gas structures, many aspects of the multiphase gas are still open questions and remain to be investigated in both small-scale simulations and subgrid modelling. Some of such aspects are:

\begin{itemize}
    \item \textit{Magnetic fields}: Although magnetic fields can suppress mixing via hydrodynamical instabilities like Kelvin-Helmholtz instabilities \citep{Chandrasekhar1961HydrodynamicStability, Ji2019SimulationsLayers}, as shown in \citet{Das2024}, the mixing properties depend only on the turbulent velocities. Hence, \model model should remain unchanged with or without magnetic fields.
    \item \textit{Thermal conduction}: In the turbulent mixing of cold gas, the eventual mixing of cold gas is via thermal conduction at molecular diffusion scales. Such small scales are not resolved in our simulations, but as shown in \citet{Tan2021RadiativeCombustion}, the turbulent mixing is rate-limited at the scales of the largest eddy, and is converged with the largest eddy scales are well resolved. Although, in our simulations, the thermal conduction is numerical, due to the rate-limiting nature of the largest scale, we expect the multiphase gas behaviour to be unchanged.
    \item \textit{Viscosity}: Similar to thermal conduction, viscosity operates at very small scales and can change the small-scale turbulent properties. Even though we have numerical viscosity in our simulations, due to the same rate-limiting nature of largest eddies in turbulent mixing, with grid cells larger than the viscous scales, the multiphase gas properties are expected to be unaffected\footnote{Note, however, that viscosity can change the microturbulent properties of the `laminar' cooling front on small scales, and thus, potentially affect the form of our coupling terms \citep[cf.][]{Tan2021RadiativeCombustion}. 
    This will need to be checked explicitly with small-scale simulations in the future (Marin-Gilabert et al., in prep.).
    .}.
\end{itemize}

Apart from the limitations mentioned above, there are many directions the \model can be expanded in future studies. Currently, \model does not account for in-situ formation of cold gas from hot gas via processes like thermal instability \citep{Field1965THERMAL1965, McCourt2018AGas, Sharma2010ThermalClusters}. This will enable the creation of the initial seed cold gas which can later grow further via turbulent mixing included in the \model model.

Another avenue for refining the model is the inclusion of other phases. It is clear from observations that colder ($<8000$K) gas exists \citep{McKee1977ASubstrate, Cox2005ARA&A..43..337C},
which points to the existence of a three-phase gas, with much more rich physics and complex interactions among the phases \citep{Farber2021TheWinds,Chen2024}. Three-phase turbulent gas is still a relatively unexplored system and detailed investigations with small-scale simulations are required before a theory can be developed to be included in \model.

While several ingredients are still missing, this work represents a first step for multi-fluid cosmological simulations. These next generation of large-scale simulations would overcome the vexing converging issue in particular in the CGM that current models suffer (see discussion in \S~\ref{sec:discu_need}). In contrast -- as demonstrated in \S~\ref{subsubsec:ver_rad_mix_grow} -- the total cold gas mass of \model is independent of resolution, and thus can lead to converged large-scale simulations and a robust comparison to observations. To do so, however, it is important to recall that multi-fluid simulations are not as easily interpretable as single-phase ones. For instance, the sub-resolution morphology and kinematics have to be defined as both absorption \citep{Hummels2017Trident:Simulations, Mukesh2024arXiv241117173S, Rudie2019ApJ...885...61R} 
as well as emission \citep[e.g.][]{GronkeLya2016ApJ...833L..26G, Hansen2006MNRAS.367..979H, Chang2024MNRAS.532.3526C}
crucially depends on them. While some results (such as the clump mass distribution; \citealp{GronkeTurb2022, Tan2023CloudWinds})
are already known, and more small-scale simulations are needed to parametrize this information. This information can then be used to alter tools such as \texttt{Trident} \citep{Hummels2017Trident:Simulations} to make them aware of the subgrid details in a multi-fluid simulation.

\section{Conclusions}
\label{sec:conc}

In this study, we introduce our new \model subgrid model to account for the subgrid cold gas behaviour, using multifluid hydrodynamics. We use the theoretical framework developed and confirmed in previous work \citep{Fielding2020MultiphaseLayers, Tan2021RadiativeCombustion, GronkeTurb2022, Das2024}, and the multifluid implementation \citep{Weinberger2023} in \texttt{AREPO} \citep{Springel2010MNRASArepo, Rainer2020AREPOPublic}. First, we present the details of our models which consists of, 
\begin{itemize}
    \item Mass, momentum and energy fluxes ($\Qdot$), from drag forces ($\Qdrag$), turbulent mixing ($\Qmix$), and cold gas growth ($\Qgrow$).
    \item A local turbulent velocity estimation methods ($v_{\rm turb}$) based on  a Kolmogorov scaling-based method ($v_{\rm turb, \texttt{kol}}$), or a velocity gradient-based approach ($v_{\rm turb, \texttt{grad}}$).
    \item an estimate of the cold gas surface and cross-sectional area  ($2h(\alpha)$ and ($A_{\rm cross}(\alpha)$, respectively).
\end{itemize}

Second, we separately verify the different parts of the \model model for the two local turbulence estimates. We compare the \model runs with resolved benchmark single-fluid \texttt{Athena++} simulations for verification. We test the quantities across different turbulent Mach numbers, resolved/unresolved initial cold clouds, resolution and random seeds for turbulent driving, for a robust comparison. The following are the main conclusions from our verification and tests in this study:
\begin{itemize}
    \item We show that the reduced version of the \model model for non-radiative mixing, i.e. $\Qdot_{\rm non-rad} = \Qdrag + \Qmix$, matches the benchmarks non-radiative \texttt{Athena++}, both qualitatively and quantitatively, in terms of the destruction timescales, across all different parameters, with very similar mean and scatter in $t_{\rm half}/t_{\rm cc}$, 
    \item We find that \model model leads to physically consistent interaction between the phases in both resolved and unresolved cold gas structures, where the mass exchange only happens at the interfaces when the cold gas is resolved and throughout the structure when unresolved, 
    \item The full \model model for radiative mixing, shows the expected behaviour of cold gas growth at short cooling timescales and cold gas destruction for long cooling timescales, across all different parameters,
    \item We verify that the full \model model for radiative mixing, quantitatively matches the expected cold gas mass growth timescales, with very similar mean and scatter in $t_{\rm grow}/t_{\rm grow, theory}$ between the benchmark \texttt{Athena++} runs and \model runs. In both, we find the mean $t_{\rm grow}$ to be $\sim 1.5 t_{\rm grow, theory}$,
    \item The full \model model for radiative mixing, also recreates the cold gas survival criterion from \citet{GronkeTurb2022}, as an emergent process, i.e., while `survival' is not explicitly implemented in \model, it can recover this larger scale result.
    \item We show that the cold gas dispersion is similar between the full \model model for radiative mixing and analogous benchmark \texttt{Athena++} runs. The agreement is better in case of a resolved, initial cold gas cloud, compared to the unresolved initial cold gas cloud,
    \item All the verification tests hold true, regardless of the local turbulent velocity estimation method,
    \item We demonstrate the strength of the \model model by running a $64^3$ cells simulation using \model with $100$ unresolved clouds, which would require $\sim 3000^3$ cells in a single-fluid code without a subgrid model.
\end{itemize}

Our study presents our new physically motivated, multifluid subgrid model \model. We have extensively tested and verified the model across a wide range of possible simulation parameters, to ensure a robust and consistent model. This work will be a useful development towards running converged large-scale simulations with subgrid prescriptions for the unresolved cold gas. However, this \model has many avenues for improvement like the inclusion of the molecular phase, in-situ cold gas formation, subgrid turbulence prescription, etc, which we hope to tackle in future work.

\section*{Acknowledgements}
We thank Aniket Bhagwat, Bo Peng, Drummond Fielding, Matthew Smith, Rajsekhar Mohapatra, Ruediger Pakmor, and Volker Springel for their helpful discussions. HD thanks staff and colleagues at the Max Planck Insitute for Astrophysics and The International Max Planck Research School on Astrophysics for their valuable support during the research.
MG thanks the Max Planck Society for support through the Max Planck Research Group.
RW acknowledges the funding of a Leibniz Junior Research Group (project number J131/2022).
Computations were performed on HPC systems Freya and Orion at the Max Planck Computing and Data Facility.

This research made use of \texttt{AREPO} \citep{Springel2010MNRASArepo, Rainer2020AREPOPublic}, \texttt{Athena++} \citep{Stone2020},\texttt{NumPy}\citep{Harris2020ArrayNumPy}, \texttt{matplotlib} \citep{Hunter2007Matplotlib:Environment}, \texttt{SciPy} \citep{Virtanen2020SciPyPython} and \texttt{inspector-gadget}.

\section*{Data Availability}
Data related to this work will be shared on reasonable request to the corresponding author.

\bibliographystyle{mnras}

\bibliography{filtered} 

\appendix

\section{Local Turbulence estimation in 2D}
\label{app:turb_2d}
We can generalise the expression for 2D geometries. In 2D, the limits of the integral are different, along with a different definition of the cell volume, leading to a slight variation in Eq.~\eqref{eq:final},
\begin{alignat}{1}
    &v_{\rm turb, \texttt{grad}, \rm 2D} = (\sigma_{v_x}^2 +\sigma_{v_y}^2 )^{1/2}
    = V_{\rm cell}^{1/2}\sqrt{\frac{1}{\xi_{\rm 2D}}\sum_{i, j}^2\left(\frac{\partial v_j}{\partial x_i}\right)^2} \label{eq:final_2D}
\end{alignat}
where, $\xi_{\rm 2D} = 4$.

During our non-radiative turbulent mixing tests, explained later in Sec.~\ref{sec:verify_local}, we find a $\xi_{3D} = 2$ works better in matching with the benchmark Athena++ simulations. Hence, we use a $\xi_{3D} = 2$ in \model runs. We can also combine both Eq.~\eqref{eq:final}~\&~\eqref{eq:final_2D} into a single expression for $D$ dimensions,
\begin{alignat}{1}
    &v_{\rm turb, \texttt{grad}, \rm D} = V_{\rm cell}^{1/D}\sqrt{\frac{1}{\xi_{\rm D}}\sum_{i, j}^D\left(\frac{\partial v_j}{\partial x_i}\right)^2} \nonumber  \\
    & \text{where, } \xi_{\rm D} =
    \begin{cases}
        4 & \text{if } D = 2 \\
        3 & \text{if } D = 3 \text{ Analytical} \\
        2 & \text{if } D = 3 \text{ \model} \\
    \end{cases}
    \label{eq:final_D}
\end{alignat}


\bsp	
\label{lastpage}
\end{document}